\documentclass[aps,prl,reprint,superscriptaddress,longbibliography]{revtex4-1}
\usepackage[english]{babel}
\usepackage{amsmath,amssymb,bbm,graphicx,color,comment}
\usepackage[bookmarks=true,colorlinks,citecolor=blue,urlcolor=blue]{hyperref}

\usepackage{dsfont}
\usepackage{soul}
\usepackage{subfigure}
\usepackage{extarrows}
\usepackage{float}

\usepackage{mathtools}
\usepackage[usenames,dvipsnames]{xcolor} 
\usepackage[export]{adjustbox}
\usepackage{adjustbox}

\usepackage{dcolumn}   
\usepackage{bm}        
\usepackage{amsfonts}  
\usepackage{filecontents}
\usepackage{lineno}

\usepackage{times}

\newcommand{\an}{\quad \quad \mathrm{and} \quad\quad}

\newcommand{\bra}[1]{\left\langle #1 \right|}
\newcommand{\exv}[1]{\left\langle #1 \right\rangle}
\newcommand{\comm}[2]{[ #1 , #2 ]}
\newcommand{\pwisein}{\left\{ \begin{array}{ll}}
	\newcommand{\pwiseout}{\end{array}\right.}
\newcommand{\ket}[1]{\left| #1 \right\rangle}
\newcommand{\bracket}[2]{\left\langle #1 | #2 \right\rangle}

\DeclareMathOperator{\tr}{\mbox{tr}}

\DeclareMathOperator{\sign}{{sign}}

\newcommand{\irrep}[1]{{\bf D}^{#1}}

\newcommand{\eqw}[1]{(\ref{#1})}
\newcommand{\eq}[1]{Eq.~(\ref{#1})}

\newcommand{\fig}[1]{Fig.\thinspace{}\ref{#1}}
\newcommand{\figg}[2]{Fig.\thinspace{}\ref{#1} and \ref{#2}}
\newcommand{\fc}[1]{({#1})}
\newcommand{\figc}[2]{Fig.\thinspace{}\ref{#1}\thinspace{}\fc{#2}}
\newcommand{\figcc}[3]{Fig.\thinspace{}\ref{#1}\thinspace{}\fc{#2} and \fc{#3}}

\newcommand{\eu}[0]{\mathrm{e}}
\newcommand{\iu}[0]{\mathrm{i}}

\begin{document}

\title{Characterizing Topological Excitations of a Long-Range Heisenberg Model with Trapped Ions}%

\author{Stefan Birnkammer}%
\affiliation{Department of Physics, Technical University of Munich, 85748 Garching, Germany}
\affiliation{Munich Center for Quantum Science and Technology (MCQST), Schellingstra{\ss}e 4, 80799 M{\"u}nchen, Germany}
\author{Annabelle Bohrdt}
\affiliation{Department of Physics, Technical University of Munich, 85748 Garching, Germany}
\affiliation{Munich Center for Quantum Science and Technology (MCQST), Schellingstra{\ss}e 4, 80799 M{\"u}nchen, Germany}
\author{Fabian Grusdt}
\affiliation{Department of Physics and Arnold Sommerfeld Center for Theoretical Physics (ASC), Ludwig-Maximilians-Universit\"at M\"unchen, Theresienstr. 37, M\"unchen D-80333, Germany}
\affiliation{Department of Physics, Technical University of Munich, 85748 Garching, Germany}
\affiliation{Munich Center for Quantum Science and Technology (MCQST), Schellingstra{\ss}e 4, 80799 M{\"u}nchen, Germany}
\author{Michael Knap}
\affiliation{Department of Physics, Technical University of Munich, 85748 Garching, Germany}
\affiliation{Munich Center for Quantum Science and Technology (MCQST), Schellingstra{\ss}e 4, 80799 M{\"u}nchen, Germany}

\begin{abstract}
	Realizing and characterizing interacting topological phases in synthetic quantum systems is a formidable challenge. Here, we 
	propose a Floquet protocol to realize the antiferromagnetic Heisenberg model with power-law decaying interactions. Based on analytical and numerical arguments, we show that this model features a quantum phase transition from a liquid to a valence bond solid that spontaneously breaks lattice translational symmetry and is reminiscent of the Majumdar-Ghosh state. The different phases can be probed dynamically by measuring the evolution of a fully dimerized state. We moreover introduce an interferometric protocol to characterize the topological excitations and the bulk topological invariants of the interacting many-body system. 
\end{abstract}

\date{\today}%
\maketitle

\textit{Introduction.---}Recent progress in realizing synthetic quantum systems has offered new opportunities for the experimental characterization and control of topological quantum phases. Topologically nontrivial band structures have been created by periodic driving~\cite{Aidelsburger,  Miyake, Jotzu2014,  Flaschner2016, Wintersperger2020}, interaction-induced chiral propagation of excitations have been studied in the few-body limit of quantum Hall states~\cite{Tai2017, Roushan2017}, symmetry protected topological (SPT) phases have been realized~\cite{Gu_2009, Pollmann_2012, Leseleuc2019, Sompet2021}, and quantum spin liquids have been explored with quantum devices~\cite{Satzinger2021, Semeghini2021}. While first steps have been laid in realizing interacting topological phases, several challenges remain, in particular concerning the characterization and control of individual topological excitations. 

\begin{figure}[!ht]
	\centering
	\includegraphics[width=.96\columnwidth]{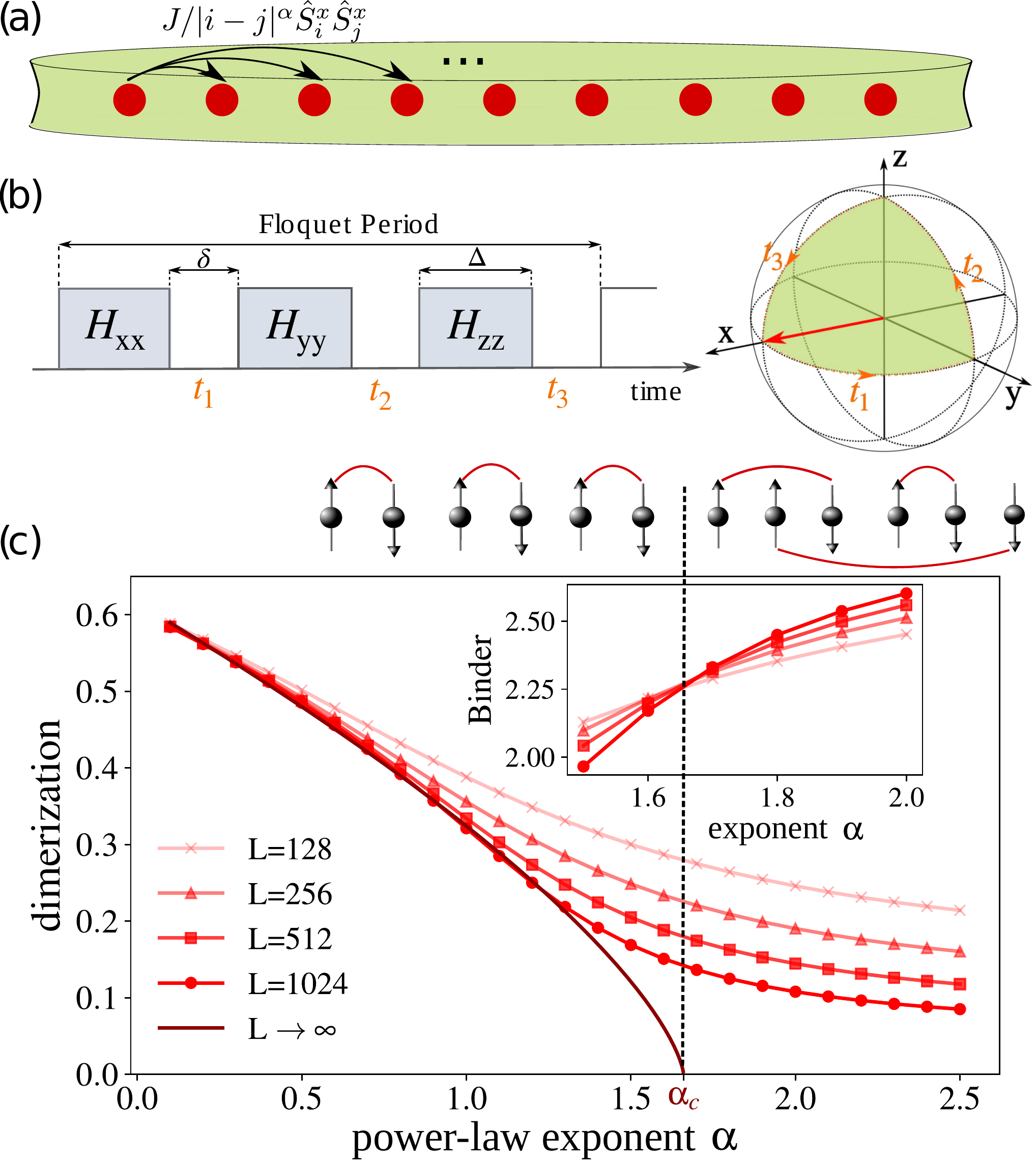}
	\caption{\textbf{Floquet protocol and phase diagram.} (a) Interactions between ions in a linear Paul trap are of Ising type $H_{xx} = J/|i-j|^\alpha \hat S^x_i \hat S^x_j$ with discrete $\mathds{Z}_2$ symmetry.  (b) Periodically applying global $\pi/2$-pulses around different axes of the Bloch sphere creates interactions along all three spin directions $H_{\text{xx}}$, $H_{\text{yy}}$, and $H_{\text{zz}}$. (c) The high-frequency limit of such a protocol realizes a long-ranged Heisenberg model $H_{\text{LR}}$ with continuous $SU(2)$ symmetry.  This model features a quantum phase transition from a dimerized to a liquid phase. Inset: A Binder cumulant analysis of the dimerization determines a critical power-law exponent of $\alpha_{c}\approx1.66$.}
	\label{fig:fig1}
\end{figure}

Here, we propose the realization of a dimerized valence bond solid with topologically nontrivial excitations in a Heisenberg model with power-law interactions using trapped ions~\cite{blatt_quantum_2012}. This phase arises due to frustration from long-range interactions and is adiabatically connected to the symmetry broken Majumdar-Ghosh phase~\cite{Majumdar1969,Majumdar1969a,White1996}. When locally deforming this Hamiltonian to introduce bond alternating couplings, it realizes a Haldane SPT phase~\cite{Haldane1983}. Our model therefore illustrates the interplay between spontaneous symmetry breaking and toplogical order.
To realize the long-range Heisenberg model in a trapped-ion setting, we propose a Floquet protocol that consists of periodic globally-applied $\pi/2$-pulses around different axes of the Bloch sphere, see \figc{fig:fig1}{a, b}. 
We determine the phase diagram and propose an interferometric protocol to characterize the topological excitations and the bulk topological invariants of our interacting many-body system. 

\textit{Model.---}We investigate a long-range spin-$1/2$ Heisenberg chain with open boundaries
\begin{equation}
	\label{eq:Long-Range-Heisenberg}
	{H}_{\text{LR}}(\alpha) \ = \   \ \sum_{i < j}  \dfrac{J}{\vert i- j \vert ^{\alpha}} \ \left[{{\hat{S}}}_{i}^x  {{\hat{S}}}_{j}^x + {{\hat{S}}}_{i}^y  {{\hat{S}}}_{j}^y + {{\hat{S}}}_{i}^z  {{\hat{S}}}_{j}^z  \right],
\end{equation}
where $\alpha$ is the power-law exponent of the long-range interactions and $J>0$ their typical energy scale (for a spin-$1$ variant of the model see Ref.~\cite{Gong_2016}). 
When considering only nearest and next-to-nearest neighbor couplings, $H_\text{LR}(\alpha)$ reduces to the Majumdar-Ghosh (MG) model~\cite{Majumdar1969,Majumdar1969a}, which exhibits a phase transition from a liquid to a dimerized valence bond solid that breaks the translational invariance of the lattice~\cite{White1996}. 

The long-range Heisenberg model $H_\text{LR}$ reduces for $\alpha \to \infty$ to the conventional Heisenberg model with nearest-neighbor couplings, whose ground state is a gapless spin liquid with power-law decaying antiferromagnetic correlations~\cite{giamarchi_quantum_2004}. For the opposite limit of $\alpha \to 0$, each spin interacts equally with all the others and the ground states correspond to arbitrary singlet pairings (see supplementary materials~\cite{supp} and references~\cite{Cornwell1998,Marshall1955, Haldane} therein). For small but finite $\alpha$, it is energetically favorable to form singlets on neighboring sites as in the Majumdar-Ghosh state
\begin{equation}
	\label{eq:MG-State}
	\ket{\text{MG}} = \prod_{i}^{L/2} (\ket{\uparrow}_{2i}\ket{\downarrow}_{2i + 1} - \ket{\downarrow}_{2i}\ket{\uparrow}_{2i + 1})/\sqrt{2}.
\end{equation} 
As a consequence the ground state breaks translational symmetry. Due to the different behavior of the ground states in the two limits,  at least one quantum phase transition occurs at some critical $\alpha_{c}$.

To quantitatively determine the phase diagram, we perform Density-Matrix Renormalization Group (DMRG) simulations~\cite{Verstraete_2008, Schollwoeck2010} and compute the dimerization order parameter
\begin{equation}
	\label{eq:dimer-operator}
	{d}_{i} \ = \ \vec{{S}}_{2i} \cdot (\vec{{S}}_{2i + 1} - \vec{{S}}_{2i - 1}).  
\end{equation}
as a function of $\alpha$ and for different system sizes $L$,  see  \figc{fig:fig1}{c}.  The data for the thermodynamic limit was extrapolated from finite-size results by a scaling Ansatz $\vert\alpha-\alpha_{c}\vert^{\nu}$. Expressions without spatial indices, indicate in the following averages over the bulk.  In our numerical studies, we represent the Hamiltonian as a matrix product operator, in which we approximate the power-law coupling by a sum of exponentials~\cite{Murg2008, Schollwoeck2010}. An analysis of the Binder cumulant $\exv{d^4}/\exv{d^2}^2$, that is expected to be scale-independent at criticality~\cite{Binder2010, Landau2015}, indicates a quantum phase transition from a dimerized phase to a liquid at $\alpha_{c} \approx 1.66$ (inset). 
Precisely extracting the critical point is a formidable challenge, as the transition is in the Berezinskii-Kosterlitz-Thouless universality class~\cite{White1996, Eggert, Eggert1996}.
We emphasize, however, that for the following discussions the precise location of the phase transition is not crucial.

\begin{figure}
	\centering
	\includegraphics[width=\columnwidth]{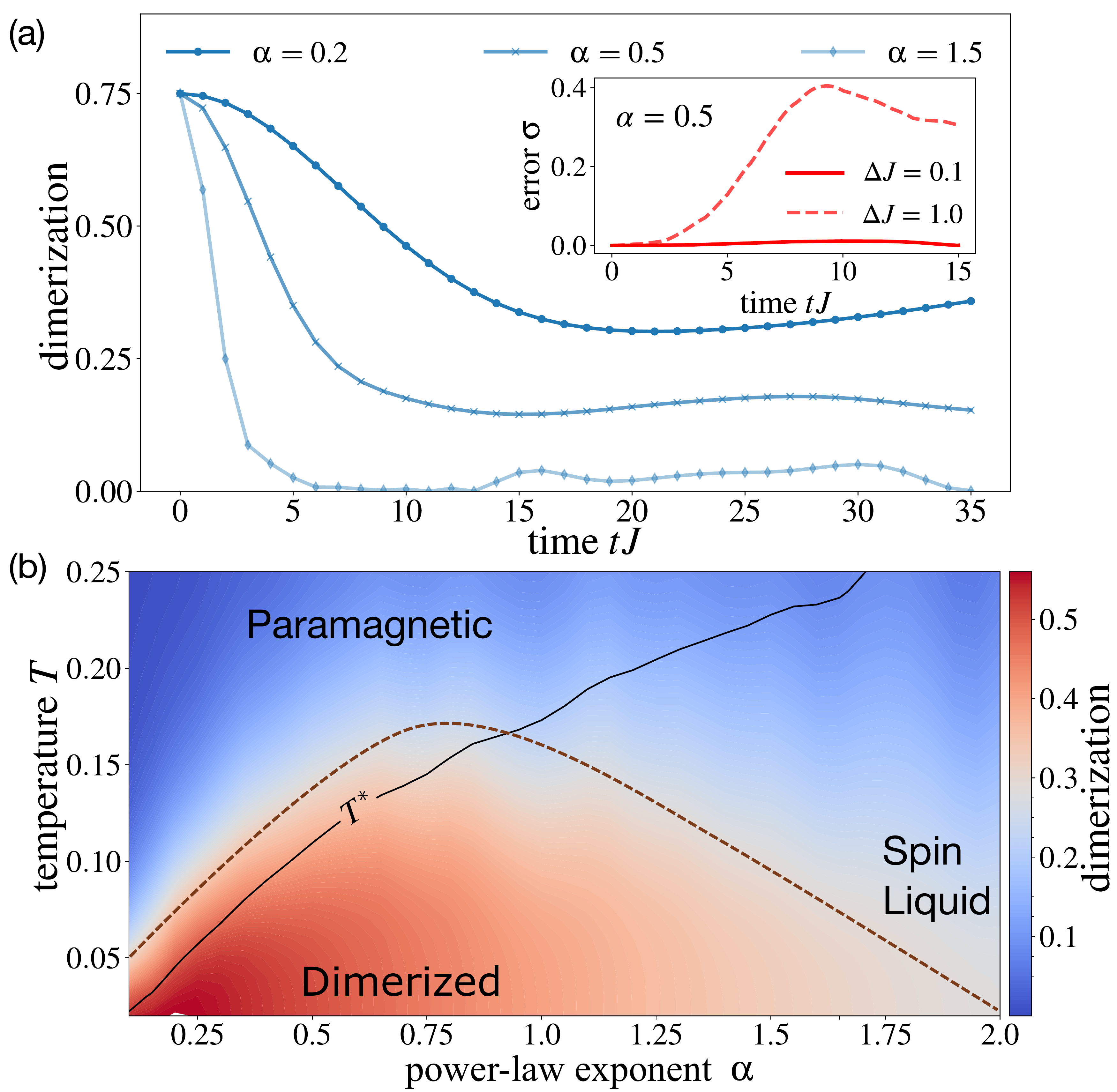}
	\caption{\textbf{Dynamical phase diagram.} (a) Evolution of the dimerization $d$ under the Floquet dynamics for an initial $\ket{\text{MG}}$-state of singlets. The evolution is governed by $H_{\text{LR}}(\alpha)$ for $\alpha\in\{0.2, 0.5, 1.5\}$ and a chain of 64 sites. Inset: We illustrate the relative deviation $\sigma\equiv\vert (d_{\Delta J\rightarrow0}-d_{\Delta J})/d_{\Delta J\rightarrow0}\vert$ of the Floquet protocol from the exact evolution for different values of $\Delta J$ and $\alpha=0.5$. (b) Thermal phase diagram for $H_{\text{LR}}(\alpha)$ including the effective temperature $T^{*}$ of our system (solid line) and a schematic of the phase boundary between the dimerized and the translational invariant phases (dashed dome) for a system of 18 sites.}
	\label{fig:dynamics}
\end{figure}

\textit{Floquet Protocol.---}Collective vibrations of an ion crystal mediate long-range Ising interactions $H_{xx} = \sum_{i<j} J/|i-j|^\alpha S^x_i S^x_j$~\cite{richerme_non-local_2014, jurcevic_quasiparticle_2014}. Previous works suggested to use multiple phonon branches~\cite{Porras_2004} and quasi-periodic driving~\cite{Bermudez_2017} to realize Heisenberg type interactions, or have employed digital simulation schemes~\cite{Lanyon_2011}. Here, instead we suggest periodic driving~\cite{Choi, Arrazola2016, Kasper2020, Agarwal2019} to promote the discrete $\mathds{Z}_2$ symmetry of the Ising interactions to the continuous $SU(2)$  symmetry of the Heisenberg interactions. The protocol consists of $\pi/2$-pulses around different axes to encircle a surface of the Bloch sphere \figc{fig:fig1}{b}; see Ref.~\cite{Arrazola2016} for a related protocol and Refs.~\cite{Geier2021, Scholl2021, Kranzl2022} for recent experimental realizations. The duration  $\delta$ of the $\pi/2$-pulses can be chosen to be much shorter than the waiting time $\Delta$, leading to an effective  period of $\approx 3 \Delta$. This way the many-body state rotates periodically from the $x$ over $y$ to $z$ direction. These unitary transformations can also be interpreted to act on the Hamiltonian instead of the many-body state, leading to an effective time evolution with alternating $H_{xx}$, $H_{yy}$, and $H_{zz}$ Ising couplings. Provided the rate $\Delta^{-1}$ is fast compared to the typical interaction strength $J$, a high-frequency expansion~\cite{Bukov2014} for the effective periodic drive can be computed, which to leading order yields the $SU(2)$ invariant Hamiltonian \eqw{eq:Long-Range-Heisenberg}. 

\textit{Singlet evolution.---}As a direct application of our Floquet protocol, we compute the time evolution of the long-range Heisenberg model $H_\text{LR}(\alpha)$ for an initial singlet state  $\ket{\text{MG}}$ using the time-dependent variational principle for MPS~\cite{Haegeman, Haegemana, zunkovic_2018, Hauschild_2018, Halimeh2016}. 
For quenches to small $\alpha$, we find that the dimerization remains finite at long times, whereas it quickly decays to zero for quenches to large $\alpha$, see \figc{fig:dynamics}{a}. 
To compare the discrete Floquet evolution with the exact dynamics of $H_\text{LR}(\alpha)$, we introduce the relative deviation $\sigma\equiv\vert (d_{\Delta J\rightarrow0}-d_{\Delta J})/d_{\Delta J\rightarrow0}\vert$, shown in the inset of \figc{fig:dynamics}{a} for $\alpha=0.5$. We find that the Floquet protocol accurately describes the $SU(2)$ invariant Heisenberg evolution for $\Delta J = 0.1$ (depending on $\alpha$, larger values of $\Delta J \sim 1$ can be safely reached~\cite{supp}). 

The energy density of the dimerized initial state is $\bra{\text{MG}}H_{\text{LR}}(\alpha)\ket{\text{MG}}/L=-0.375J$ independent of $\alpha$, which is larger than the ground-state energy density of $H_\text{LR}(\alpha)$. Hence, the quench deposits an extensive amount of energy into the system. 
According to the eigenstate thermalization hypothesis~\cite{deutsch_quantum_1991, srednicki_chaos_1994, rigol_thermalization_2008}, which is expected to hold for generic interacting systems as this, a subsystem should thermalize to an effective temperature $T^*$, that is consistent with the energy density deposited in the system. The effective temperature can then be evaluated self-consistently from the condition $\bra{\text{MG}}H_{\text{LR}}(\alpha)\ket{\text{MG}} =  \tr[ H_{\text{LR}}(\alpha) {e}^{-H_{\text{LR}}(\alpha)/T^{*}}/\mathcal{Z}]$ where $\mathcal{Z}$ is the partition sum.
We approximate the thermal expectation value using the typicality approach~\cite{Bartsch2009} and evolve 50 random initial states in imaginary time using exact diagonalization on 18 spins (see \cite{supp} for system size dependence) to extract effective temperature $T^*$ as a function of $\alpha$; \figc{fig:dynamics}{b}. 
From the dynamical phase diagram, we find that for $\alpha \lesssim 1$ the effective temperature is low enough such that a finite dimerization is supported in the steady state, whereas it decays to zero for $\alpha \gtrsim 1$, consistent with our observations on the time evolution in \figc{fig:dynamics}{a}.  
Despite the one-dimensional character of our system, a sharp finite-temperature phase transition can arise in the thermodynamic limit because of the long-range interactions~\cite{Auerbach1994}.

\textit{Measuring the Zak phase.---}Starting from the $\ket{\text{MG}}$-state, a state close to the ground state of $H_\text{LR}(\alpha)$ can be prepared by adiabatically tuning $\alpha$ as long as the system remains in the dimerized phase. We will now present a protocol to measure a topological order parameter of such a state. Following Refs.~\cite{Hatsugai2006, Hatsugai2006a}, we introduce the  $SU(2)$-transformation 
\begin{equation}
	\label{eq:Gaugetransformation}
	\Phi : \vec{S}_{i}\cdot \vec{S}_{j} \ \mapsto \ \hat{S}_{i}^{\mathrm{z}}\hat{S}_{j}^{\mathrm{z}} + \dfrac{1}{2} (\mathrm{e}^{\mathrm{i}\varphi} \hat{S}_{i}^{+}\hat{S}_{j}^{-} + \text{h.c.}),
\end{equation}
where $\varphi$ is a compact variable in the interval $[0, 2\pi]$. The $SU(2)$-transformation is designed such that it affects only couplings crossing the $\ell$-th bond between sites $\ell$ and  $\ell+1$, which separates our system into a left $S_L$ and a right $S_R$ part. All interactions within one subsystem remain unchanged. As a consequence we obtain for every choice of the bond $\ell$ a new family of Hamiltonians $H_{\text{LR}}(\alpha;\varphi)$ parametrized by $\varphi \in [0, 2\pi]$.

\begin{figure}
	\centering
	\includegraphics[width=\columnwidth]{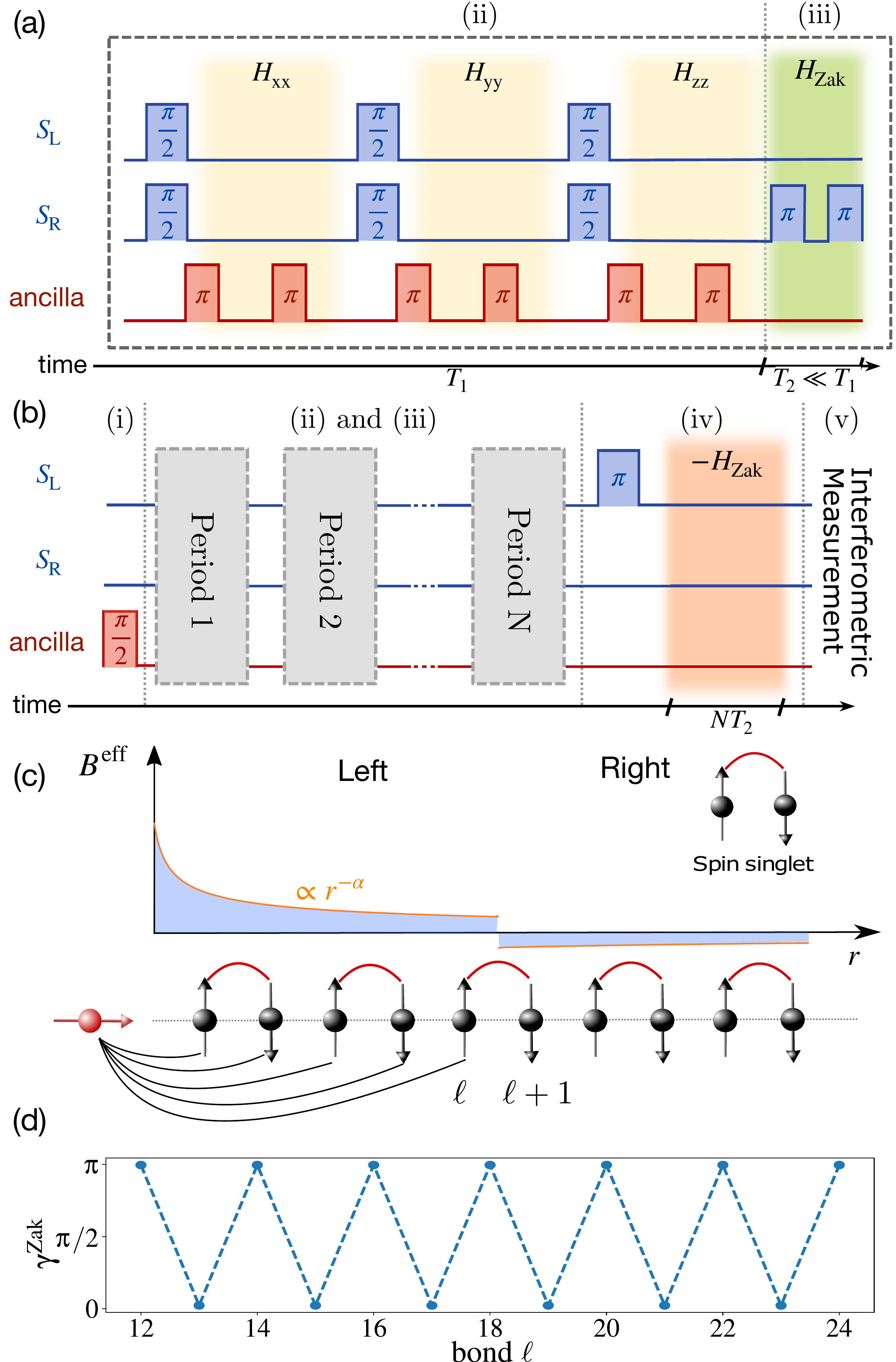}
	\caption{\textbf{Measuring many-body Zak phases with interferometry.} (a) Single Floquet period and (b) complete pulse sequence for extracting the Zak phase; see main text for details. (c) The effective field $B^\text{eff}$ acting on the system during the step (iii) of the protocol has a sign change at the bond where we measure $\gamma^\text{Zak}_\ell$. (d) Performing measurements for different $\ell$ yields alternating Zak phases of $0$ or $\pi$. Data is evaluated for a system of $32$ sites and $\alpha = 0.2$.}
	\label{fig:ZakPhase}
\end{figure}

A key for obtaining a  quantized topological order parameter is that the chosen parametrization retains the time-reversal symmetry of $H_{\text{LR}}(\alpha;\varphi)$~\cite{Hatsugai2006}.
Tuning $\varphi$ continuously through the interval $[0,2\pi]$, describes a closed loop $\mathcal{C}_{\ell}$ within the set of Hamiltonians. This allows us to introduce the Zak phase 
\begin{equation}
	\label{eq:ZakPhase}
	\gamma^{\mathrm{Zak}}_\ell = \oint_{\mathcal{C}_\ell} \mathrm{d}\varphi \bra{\psi(\varphi)}\mathrm{i}\partial_{\varphi}\ket{\psi(\varphi)},
\end{equation}
where $\ket{\psi(\varphi)}$ is the ground state of $H_{\text{LR}}(\alpha;\varphi)$. The Zak phase is well defined, provided the corresponding path $C_\ell$ is followed adiabatically, which can be ensured because the dimerized phase is gapped, see supplementary material~\cite{supp}. In order to gain some intuition about the Zak phase, we first apply the $SU(2)$-transformation to the fully dimerized $\ket{\text{MG}}$-state. When the bond $\ell$ lies within a singlet, the transformation gives  $(\ket{\uparrow\downarrow} - \mathrm{e}^{\mathrm{i}\varphi}\ket{\downarrow\uparrow})/ \sqrt{2}$.
Evaluating \eq{eq:ZakPhase} for this particular case reveals a Zak-phase of $\pi$~\cite{Grusdt2019, Grusdt2014}, while it is zero when the bond $\ell$ lies between two singlets. For the $\ket{\text{MG}}$-state and thus also for the adiabatically connected ordered ground states of $H_\text{LR}$ we consequently expect to find a Zak phase alternating between values of $0$ or $\pi$ when traversing the bond $\ell$ through the system.

Before we numerically compute the Zak phase of the dimerized state, we introduce a protocol to experimentally measure it in a trapped ion setting. Let us first gain some intuition: To realize a transformation similarly to~\eqref{eq:Gaugetransformation}, we can use an effective (in general time-dependent) magnetic field $B_i^\text{eff}(t)$ acting on the spins of $H_\text{LR}(\alpha)$, that is proportional to a step function with the step being located at bond $\ell$. Using the Peierls substitution, the magnetic field can be absorbed into Hamiltonian as $H_\text{LR}(\alpha; \varphi(t)) = \sum_{i < j} J/|i-j|^\alpha [ S^{\mathrm{z}}_{i}S^{\mathrm{z}}_{j} + \frac{1}{2} ( \mathrm{e}^{\mathrm{i}\varphi_{ij}(t)} S^{+}_{i}S^{-}_{j} + \text{h.c.} )  ]$, where $\varphi_{ij}(t) \equiv \int_{0}^{t}\mathrm{d}t' [B^\text{eff}_{j}(t') - B^\text{eff}_{i}(t')]$. A phase is only picked up, when bond $\ell$ is crossed as $B^\text{eff}_i$ is assumed to be constant except across bond $\ell$. The time $t$ is chosen such that the phase $\varphi$ is adiabatically tuned from 0 to $2\pi$. The Zak phase can be measured using a Ramsey sequence to cancel dynamical phases~\cite{Atala2013, Duca2015, Abanin}.

In order to implement this approach in a chain of ions, we identify the left-most ion as an ancilla qubit $\tau^z$ that operates on the same computational basis and has the same power-law coupling to the other spins of the chain $\sum_i J/|i|^\alpha \tau^z \hat{S}_i^z \equiv \sum_i B_i^\text{eff} \hat{S}_i^z$. The protocol then consists of the following steps, see \figcc{fig:ZakPhase}{a}{b} for an illustration: (i)  After ground state preparation of the chain initialize the ancilla qubit in a superposition state by applying a $\pi/2$-rotation. This leads to an opposite sign in $B^\text{eff}$ for the two ancilla states and in turn allows for a cancellation of the dynamical phase. (ii) Perform global $\pi/2$-rotations on the system around different axes, as discussed in \figc{fig:fig1}{c}, to realize the long-range Heisenberg dynamics. During that time perform equally spaced $\pi$-rotations on the ancilla qubit to cancel the phase accumulation from interaction with the remaining chain. (iii) After a Floquet period, apply a $\pi$-rotation only on the right part of the system to create a kink in the effective field $B_i^\text{eff}$ at bond $\ell$, see \figc{fig:ZakPhase}{b}. Accumulate phase on the ancilla and apply another $\pi$-rotation to restore the couplings within the system. Steps (ii) and (iii) are then repeated until $\varphi$ covers the whole interval $[0,2\pi]$.
(iv) Apply a $\pi-$pulse to the left part of the system enabling an inverse rotation to compensate the effect of the protocol on the wave function of the system. (v) Measure the phase of the ancilla, which corresponds to the many-body Zak phase at bond $\ell$. For a more detailed description see supplementary material~\cite{supp}

We now numerically evaluate the Zak phase \eq{eq:ZakPhase}. Using DMRG we compute the ground state for 20 discretized steps along $\mathcal{C}_\ell$ for a system of 32 sites and $\alpha=0.2$, see \figc{fig:ZakPhase}{d}, which confirms that the Zak phase is alternating between 0 and $\pi$, a characteristics of a dimerized state. 
We also confirm the adiabaticity of the computation by calculating the product of projectors into the ground state $\mathcal{P}(\varphi) = \ket{\psi(\varphi)}\bra{\psi(\varphi)}$ during each step of the protocol, which attains large values between 0.75 to 1. 

In the proposed experimental protocol, the effective magnetic field has the required jump at bond $\ell$ to introduce the local $SU(2)$-transformation, but also varies slowly across the other bonds. By numerically simulating the protocol of \fig{fig:ZakPhase}, we find that the phase accumulated on the ancilla has the characteristic bond alternating pattern for dimerized states; see supplementary material~\cite{supp} for details, where we also analyze the number of required operations.

\begin{figure}
	\centering
	\includegraphics[width=\columnwidth]{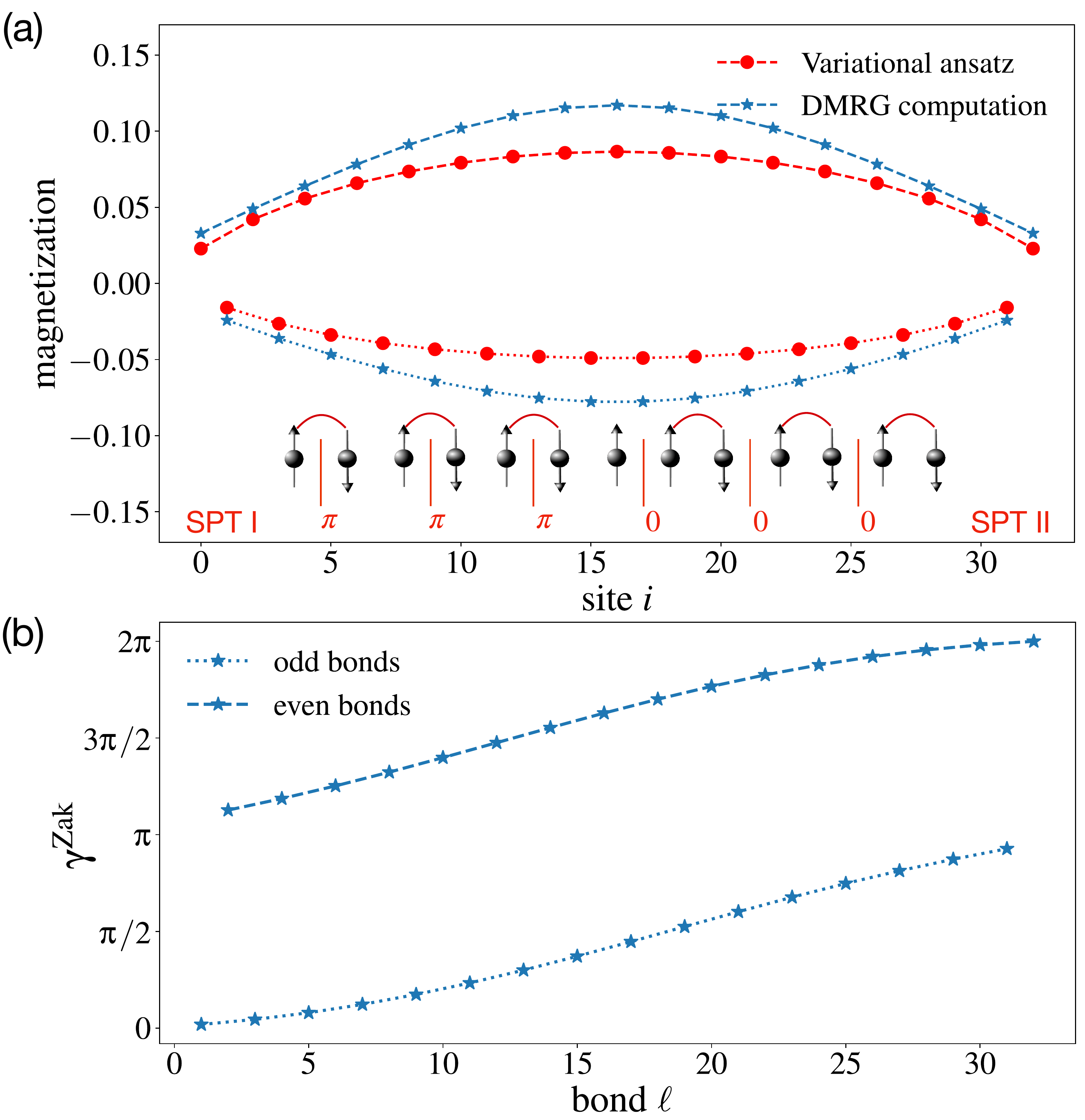}
	\caption{\textbf{Topological excitations.} (a) Local magnetization for a spinon in a lattice of 33 sites and $\alpha=0.2$. We compare the DMRG result with a variational Ansatz delocalizing a single spinon. (b) This bulk excitation swaps the Zak phase of even and odd bonds as the bond $\ell$ traverses the system.}
	\label{fig:Excitations}
\end{figure}

\textit{Topological excitations.---}In order to characterize the topological excitations of the dimerized phase, we now consider a chain with an odd number of sites. In this case, singlets cannot fully cover the chain, and hence an unpaired spin-$1/2$ (spinon) excitation is always present. Due to prominent examples such as the Affleck-Kennedy-Lieb-Tasaki (AKLT) model~\cite{Affleck1987} we are used to the existence of spin-$1/2$ states in interacting topological phases. These degenerate modes provide a clear signature for topological order and are typically localized at the edges of the system. In contrast to the usual edge modes, a measurement of the magnetization for $H_\text{LR}$ indicates that the excitation is delocalized over the entire lattice, see \figc{fig:Excitations}{a}. In order to obtain an analytical understanding, we introduce a variational state
that describes a delocalized spinon with wavevector $q$ 
separating two $\ket{\text{MG}}$-states with singlets on even and odd bonds, respectively~\cite{supp}. The spinon hence represents a defect in the topological order; inset of \figc{fig:Excitations}{a}. Variationally optimizing the ground state energy with our Ansatz yields $q=\frac{\pi L}{2(L+1)}$, which is consistent with the oscillatory magnetization pattern in \figc{fig:Excitations}{a}.  

We also characterize the Zak phase for this state. For odd numbers of sites, the dimerized ground state of the system is two-fold degenerate due to the $SU(2)$ symmetry of our model and hence we cannot construct an adiabatic path $\mathcal{C}_\ell$. To lift the degeneracy, we apply a weak local magnetic field in the center of the system, which introduces a small gap. This magnetic field breaks time reversal symmetry. From the arguments of Hatsugai~\cite{Hatsugai2006,Hatsugai2006a} it then follows that the Zak phase is not quantized in general. As a consequence, we expect a monotonous change in the Zak phase as we traverse the system which can be interpreted as a mobile domain wall separating two distinct topological phases. This is consistent with our results in  \figc{fig:Excitations}{b}, which also show that the Zak phase of even and odd bonds differ by $ \pi$ as advocated by the domain wall picture.

\textit{Outlook.---}The Heisenberg model with long-range antiferromagnetic interactions exhibits a phase transition from a liquid to a dimerized valence bond solid that spontaneously breaks the lattice translational invariance. We propose Floquet protocols for trapped ions to realize this model and to characterize the nature of the delocalized topological excitations as well as the bulk topological invariants. 

For future studies, it would be interesting to introduce an easy-axis anisotropy by adjusting the Floquet periods between the $\pi/2$-pulses to realize a deconfined quantum critical point between a dimerized and a N\'eel ordered phase in our one-dimensional model~\cite{Mudry_2019, Roberts_2019} or to explicitly break the translational symmetry by introducing bond-alternating couplings to realize a Haldane symmetry protected topological (SPT) phase with localized edge states~\cite{Haldane1983}. A future challenge is to create interacting higher-dimensional topologically ordered many-body states with synthetic quantum matter, that are characterized by a topological entanglement entropy and fractional excitations~\cite{Kiteaev_2006}.  With our protocols,  interactions in two dimensional triangular lattices of trapped ions~\cite{Bohnet2015},  could be promoted from $\mathbb{Z}_{2}$ to $SU(2)$ symmetry, with the prospect of realizing exotic frustrated mangetic states or even quantum spin liquids.  

\textit{Acknowledgements.---}We thank R. Blatt, E. Demler, Ch. Maier, F. Pollmann, and R. Verresen for insightful discussions. Our tensor-network simulations were performed using the TeNPy Library~\cite{Hauschild_2018}. We acknowledge support from the Deutsche Forschungsgemeinschaft (DFG, German Research Foundation) under Germanys Excellence Strategy--EXC--2111--390814868, TRR80 and DFG grant No. KN1254/2-1, No. KN1254/1-2, Research Unit FOR 2414 under project number 277974659, and from the European Research Council (ERC) under the European Unions Horizon 2020 research and innovation programme (grant agreement No. 851161), as well as the Munich Quantum Valley, which is supported by the Bavarian state government with funds from the Hightech Agenda Bayern Plus. \\

\begin{center}
	\textbf{\large Supplemental Materials}
\end{center}

\textit{All-to-all Interactions.---}Here, we discuss the limiting case $\alpha=0$ of $H_\text{LR}$, in which our one-dimensional model describes a cluster of spins forming one self-interacting spin of extensive magnitude $\vec{{\mathcal{S}}}$:
\begin{align}
	\label{eq:All-to-all limit}
	\hat{H}_{\text{LR}}(\alpha=0) &= J \sum_{i<j} \vec{{S}}_{i}\cdot  \vec{{S}}_{j} \nonumber \\ 
	&= \dfrac{J}{2}  \sum_{i,j} \vec{{S}}_{i} \cdot \vec{{S}}_{j}  - \dfrac{J}{2} \sum_{i} \vec{{S}}_{i}^2 \\
	&= \dfrac{J}{2} \vec{{\mathcal{S}}} \cdot \vec{{\mathcal{S}}} - \dfrac{3J}{8}L . \nonumber
\end{align}
To arrive at the final expression we introduced $\vec{{\mathcal{S}}} \equiv \sum_{i}\vec{{S}}_{i}$ and made use of the identity for amplitudes of spin operators $\vec{{S}}\cdot \vec{{S}} = s(s+1)$.

The energy spectrum of the model can now be completely determined by representation theory of the group $SU(2)$. To this end, we look at all possible irreducible representations arising from a product of $L$ spin-$1/2$ degrees of freedom. We will first assume an even number of spins $L$. The arising representations are given by\
\begin{equation}
	\label{eq:Spin-Representations}
	\bigotimes_{i = 1}^{L} ~\irrep{1/2}= \bigoplus_{j = 0}^{L/2} ~\bigoplus_{n = 1}^{N_{j}}~\irrep{j},
\end{equation} 
where we denoted an irreducible representation of spin $j$ as $\irrep{j}$ following the notation of~\citet{Cornwell1998}. Expression~\eqref{eq:Spin-Representations} indicates that we have $N_{j}$ distinct irreducible $SU(2)$-representations $\irrep{j}$, where j ranges up to total spin $L/2$. The energy $E_{j}$ corresponding to each spin representation $\irrep{j}$ is according to Hamiltonian~\eqref{eq:All-to-all limit} given as
\begin{equation}
	\label{eq:Energy Spin Representations}
	E_{j} = \dfrac{J}{2} j(j+1) - \text{const}.
\end{equation}
The ground states of the theory are hence given by all singlet representations arising from the product of $L$ spins, which is also conjectured by the theorem of Marshall for quantum antiferromagnetic Heisenberg models~\cite{Marshall1955, Auerbach1994}. 
Next we determine the degeneracies of energy levels, which are directly related to the set $\{N_{j}\}$, because we find precisely $N_{j}(2j + 1)$ states of energy $E_{j}$. First, we note that these degeneracies carry a dependence on the system size $L$. This is required to ensure an exponential growth of the Hilbert space as we extend our system with additional spins. Although also higher spin representations will be made accessible by additional spins, latter only cause a linear increase in the dimension of the Hilbert space.

\begin{figure}
	\includegraphics[width=\columnwidth]{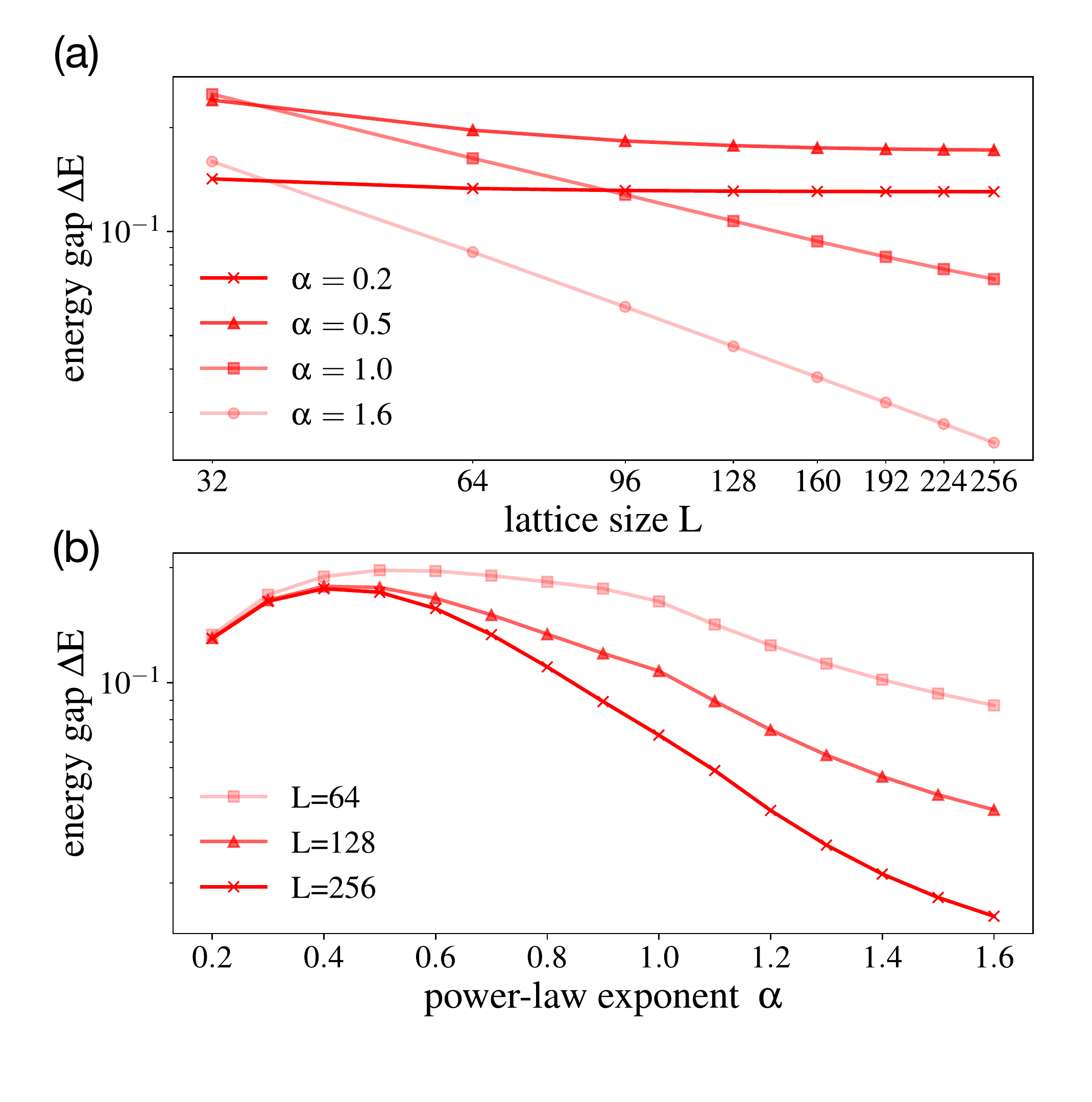}
	\caption{\textbf{Energy Gaps.} Energy gap $\Delta$E above the ground state \fc{a} as a function of the lattice size $L$ and \fc{b} as a function of $\alpha$. }
	\label{fig:Energy gaps}
\end{figure}

We can now use the $L$-dependence to determine recursive relations for every $N_{j}(L)$ respectively. For this purpose we will first look at a small system and then generalize our observations to arbitrary system sizes. We start by adding an additional spin doublet to a system of only $L=2$ spins. The corresponding decomposition into irreducible representations $\irrep{j}$ is given by
\begin{align}
	\label{eq:Spin-Rep decomposition L=4}
	\big(\irrep{1/2} \otimes  \irrep{1/2} \big) &\otimes \big( \irrep{1/2} \otimes  \irrep{1/2}  \big) \\
	&= \big( \irrep{0} \oplus \irrep{1} \big) \otimes \big( \irrep{0} \oplus  \irrep{1}  \big) \nonumber \\
	&= \big( \irrep{0} \oplus  \irrep{1}  \big) \oplus \big( \irrep{1} \oplus \irrep{0} \oplus \irrep{1} \oplus \irrep{2} \big) \nonumber\\
	&= ~~\irrep{0} \oplus \irrep{0} \oplus \irrep{1} \oplus \irrep{1} \oplus \irrep{1} \oplus \irrep{2}.  \nonumber 
\end{align}
Here, we used distributivity as well as the Clebsch-Gordan series for products of two irreducible $SU(2)$-representations.

We notice that the added singlet representation reproduces all given representations of the previous system. The triplet contained in the added spin doublet will produce three new representations for every previous one. For the example of~\eqref{eq:Spin-Rep decomposition L=4} this concretely increases $N_{0}$ by $1$, $N_{1}$ by $2$ and $N_{2}$ by $1$. If we look at the situation, given in~\eqref{eq:Spin-Rep decomposition L=4}, on more general grounds we recognize that $\irrep{0}$ can only result from a product of two singlets or two triplets. All other spin representations can, however, be generated from previous ones in four different ways. The spin can on the one hand be kept constant by an product with $\irrep{0}$ or $\irrep{1}$. On the other hand it can be increased or decreased by $1$ using the product with $\irrep{1}$. This behavior is summed up to the following recursive sequence for the set of $\{ N_{j}(L)\}$.
\begin{align}
	\label{eq:Degeneracies Energy levels}
	N_{0}(L + 2) &= N_{0}(L) + N_{1}(L) \nonumber\\
	N_{1}(L + 2) &= N_{0}(L) + 2N_{1}(L) + N_{2}(L)  \nonumber\\
	N_{2}(L + 2) &= N_{1}(L) + 2N_{2}(L) + N_{3}(L) \\
	&~~\vdots			 \nonumber \\
	N_{n}(L + 2) &= N_{n-1}(L) + 2N_{n}(L) + N_{n+1}(L) \nonumber 
\end{align} 
This set of equations can be implemented and solved symbolically, determining the entire spectrum of $H_{\text{LR}}(\alpha=0)$.

After investigating the exact limit $\alpha=0$ let us now discuss the implications for small but positive $\alpha$. Recalling Marshall's theorem~\cite{Marshall1955} we still expect to find an overall singlet as the ground state of our system. On the contrary to the previous discussion we do, however, not expect to find a degeneracy in the ground state. High degeneracies are expected for the case $\alpha=0$ as interactions do not introduce a definite order in the system and we can reorder the spins in all possible ways. This is no longer possible for $\alpha>0$, where the decaying interactions favor nearby singlets. For this reason we expect the ground state to exhibit a large overlap with $\ket{\mathrm{MG}}$, defined in the main text and possesses a finite dimerization $d$.

\begin{figure}
	\includegraphics[width=\columnwidth]{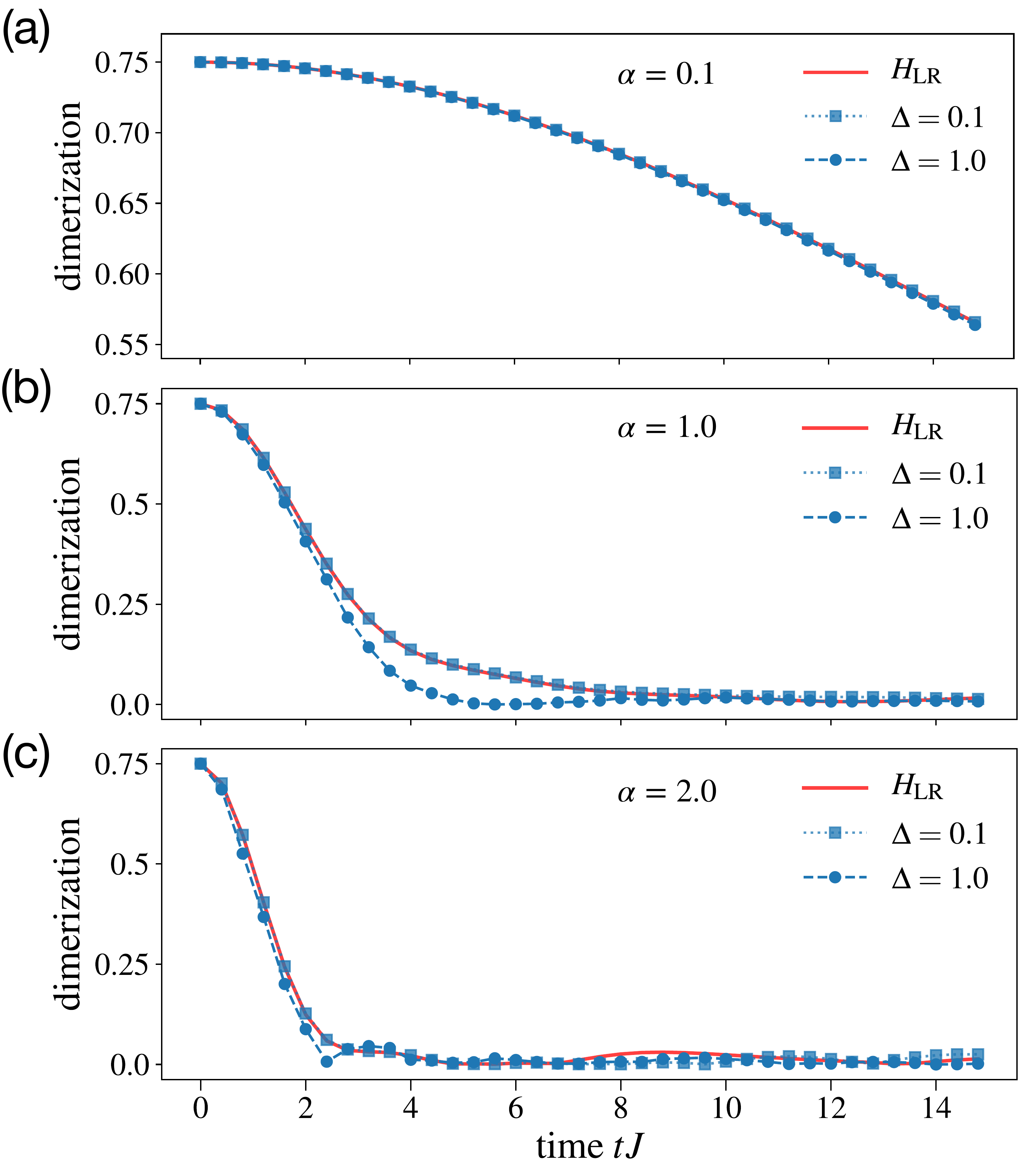}
	\caption{\textbf{Convergence of the Floquet Protocol.} The convergence of the Floquet Hamiltonian towards its infinite-frequency limit $H_{\text{LR}}$ is examined by comparing time-evolution of $\ket{\mathrm{MG}}$-states in different regions of the phase diagram: (a) $\alpha = 0.1$, \fc{b} $\alpha=1.0$, and \fc{c} $\alpha = 2.0$ for different values of the Floquet period $3\Delta$.}
	\label{fig:ConvergenceFloquet}
\end{figure}

\textit{Energy Gaps of Ordered Phase.---}After investigating the ground state of the system, we can use our previous results to discuss also excited states and especially the existence of finite energy gaps in the dimerized phase. The latter are especially important for an adiabatic implementation of our protocol to measure geometric Zak phases.

For the case of even $L$ and very small $\alpha$ we expect to find an overall spin-singlet even for the first excited states, since all spin-$0$ configurations only receive small corrections increasing with $\alpha$ compared to the degenerate case of $\alpha=0$. The gaps of all higher spin representations are, however, still determined up to a factor by~\eqref{eq:Energy Spin Representations} to be finite and $\mathcal{O}(J)$. Even though an analytical calculation of the gap is challenging, intuition can be gained from mapping our long-ranged spin model $H_{\text{LR}}$ to a Non-Linear Sigma model (NL$\sigma$M) with a topological term~\cite{Bermudez_2017,Haldane,Haldane1983}. A renormalization analysis indicates that such a system either flows to a gapless fixed point of a $SU(2)_{1}$ Wess-Zumino-Witten conformal field theory or approaches a gapped fixed point. These limits are related to the disordered gapless and the gapped ordered phase of our model. We test this conjecture numerically in \fig{fig:Energy gaps}. For small $\alpha$ our finite size analysis supports a converged energy gap above the ground state. As the critical point $\alpha_c \approx 1.66$ is approached, the required system sizes to observe an onset of saturation increases. 

A short remark should be addressed to the case of odd-numbered system sizes. Whereas finite energy gaps above the ground state were shown to exist naturally in the ordered phase for systems with an even number of sites, the situation is different for odd $L$. Although the previous arguments for a gap between the various spin representations apply in the same way for odd $L$, they also conserve the degeneracy within each representation. This fact directly results from the $SU(2)$ symmetry of $H_{\text{LR}}$. Contrary to the spin singlet for even lattice sizes transforming trivially under application of group generators from $su(2)$, the ground state spin-$1/2$-representation for odd $L$ allows a mixing of both polarization directions of the spinon excitation. The ground state degeneracy will hence remain two-fold unless we break the underlying $SU(2)$ symmetry. This can for instance be achieved with a small but finite local magnetic field, causing an energy splitting within each spin representation and creating a finite gap above the ground state.	

\textit{Convergence of the Floquet Protocol.---}We numerically study the convergence of the driven Hamiltonian to its high-frequency limit $H_{\text{LR}}$. To this end, we investigate the time-evolution of a maximally dimerized $\ket{\mathrm{MG}}$-state for both cases. The numerical results are shown in \fig{fig:ConvergenceFloquet} for Floquet step durations of $\Delta\in\{0.1,1.0\}$ and $\alpha \in \{0.1, 1, 2\}$. Note that for the following investigations we explicitly set the overall coupling strength to $J=1$, causing time intervals like $\Delta$ to be dimensionless quantities. We find fast convergence to the evolution governed by $H_{\text{LR}}$ even for comparatively low driving frequencies ($\Delta=1.0$) both deep in the dimerized phase ($\alpha=0.1$) and in the fluid phase ($\alpha=2.0$), whereas closer to the critical regime shorter periods are required. All in all, the effective Floquet Hamiltonian approximates $H_{\text{LR}}$ reasonably well already for not-too-high driving frequencies which simplifies an experimental realization.

\begin{figure}
	\includegraphics[width=\columnwidth]{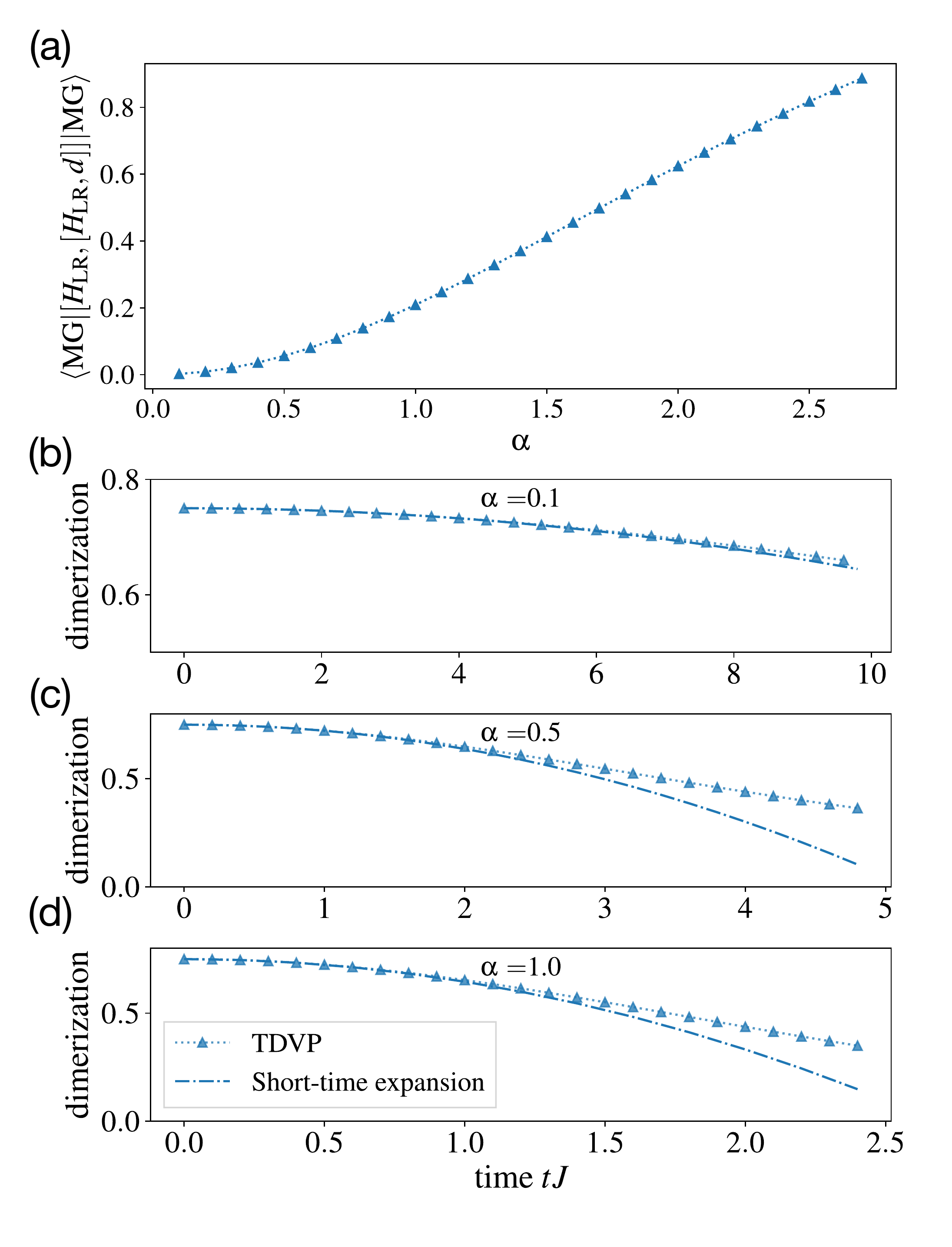}
	\caption{\textbf{Short-time evolution.} (a) We provide numerical results for the double commutator $\bra{\mathrm{MG}}\comm{H_{\mathrm{LR}}}{\comm{H_{\mathrm{LR}}}{d}}\ket{\mathrm{MG}}$ characterizing the decay of the order parameter obtained from a short-time expansion. (b)-(d) Comparison of the perturbative to the exact the time-evolution of the order parameter for values of the scaling parameter $\alpha \in \{0.1, 0.5,1.0\}$.}
	\label{fig:Figure7-Appendix}
\end{figure}

\begin{figure}
	\includegraphics[width=\columnwidth]{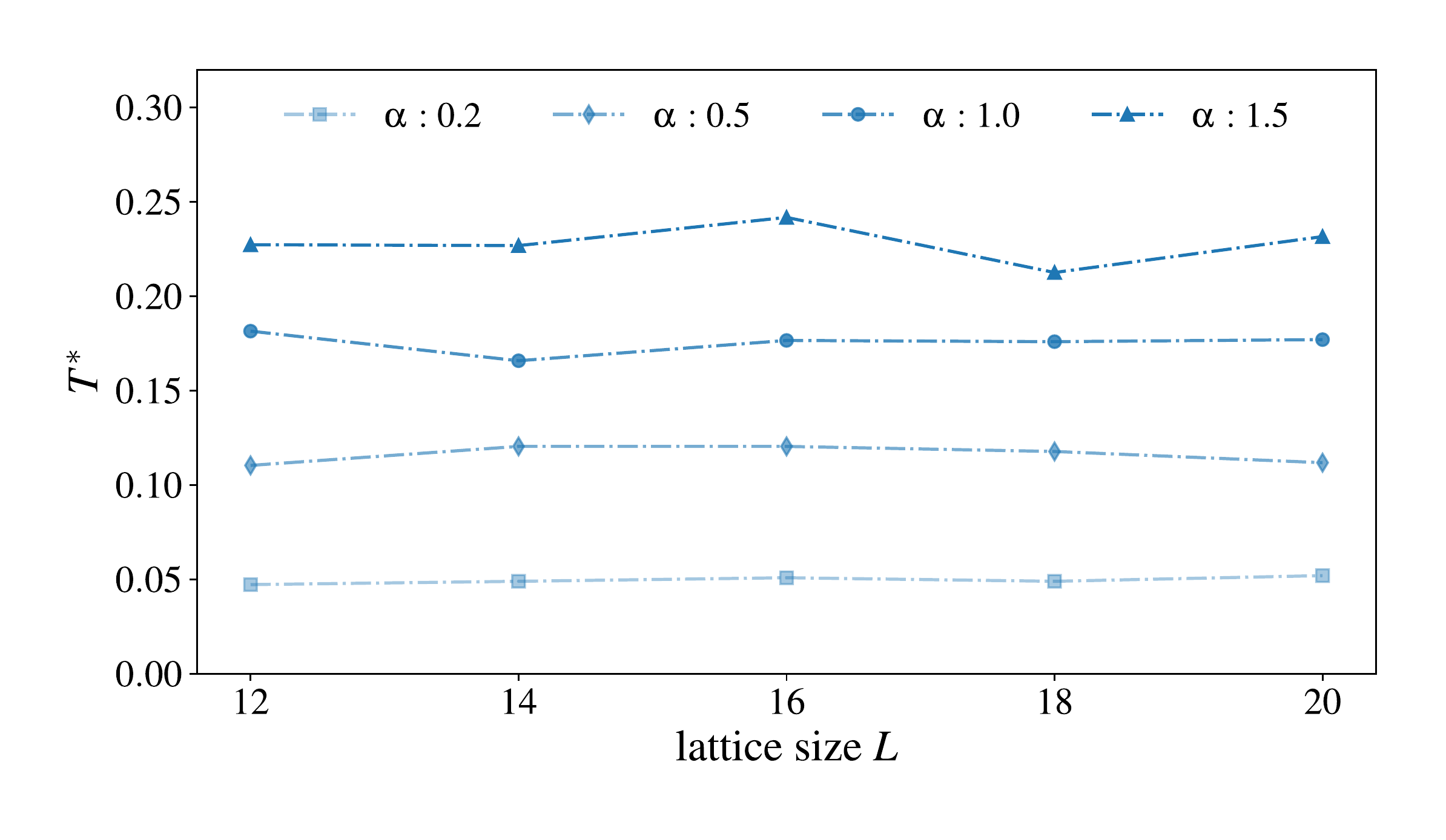}
	\caption{\textbf{Size-dependence of effective temperature.} Effective temperature $T^{*}$, at which the thermal energy of the system reaches the energy of a fully dimerized initial state, displayed for values of the scaling parameter $\alpha\in \{0.2,0.5,1.0,1.5\}$. The estimates were gathered using the typicality approach with 40 random states and system sizes from 12 to 20 sites.}
	\label{fig:Figure8-Appendix}
\end{figure}

\textit{Short-time expansion.---}Besides questions on the convergence of our protocol, a point of experimental interest should be in timescales. To estimate the typical relaxation time scale of the dimerization, we perform a short-time expansion of the quantum evolution 
\begin{figure*}
	\includegraphics[width=\textwidth]{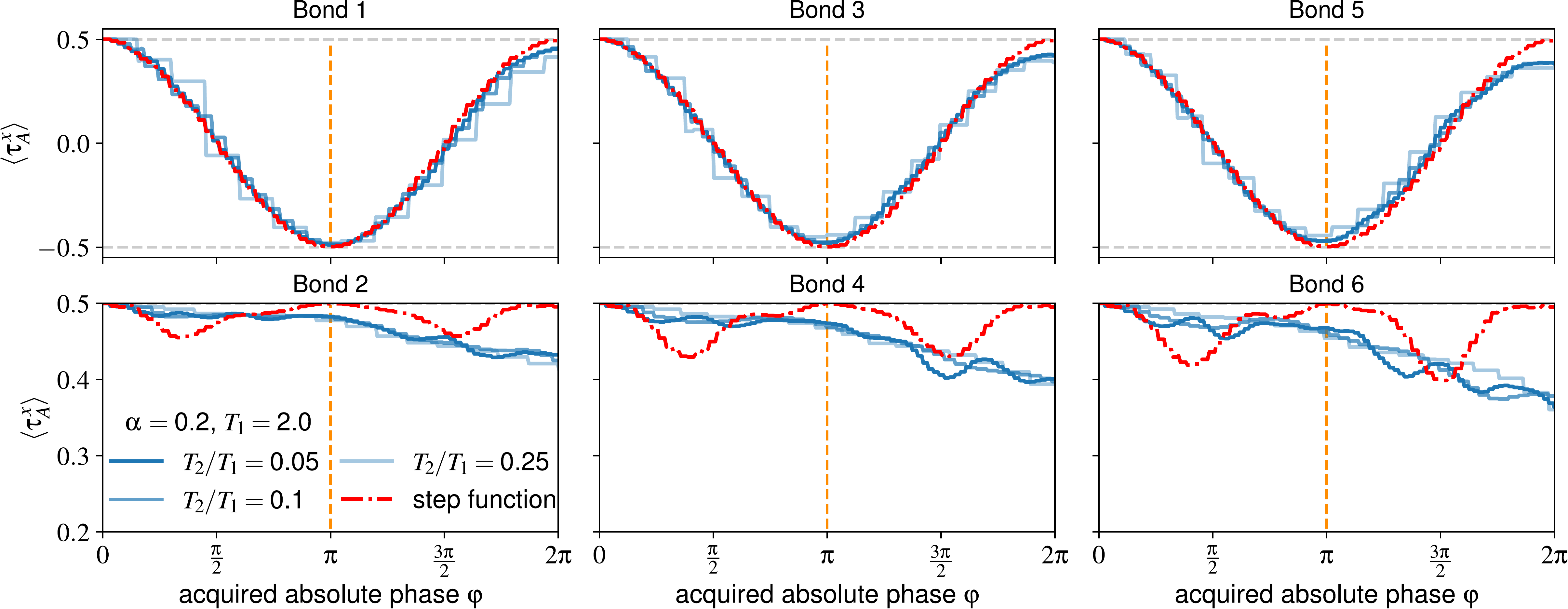}
	\caption{\textbf{Simulation of the Zak protocol.} We show the phase accumulation of the ancilla qubit  $\exv{\tau^{x}_{A}}$ during the Zak protocol illustrated in \figcc{fig:ZakPhase}{a}{b}  in the main text. We consider an effective evolution with $H_{\mathrm{LR}}(\alpha)$ for a time span of $T_{1}=2.0$ and $\alpha=0.2$. 
	The phase $\varphi$ is introduced during the evolution via application of  $H_{\mathrm{Zak}}$ for durations of $T_{2}\in\{0.1, 0.2, 0.5\}$. 
	Interferometric measurements should be performed at $\varphi=\pi$ to extract the relevant phase accumulation (orange dashed lines). The results are compared with an ideal step function coupling the ancilla and the system with $T_{2}/T_{1}=0.05$ (red dashed line).}
	\label{fig:Figure10-Appendix}
\end{figure*}
\begin{gather}
	\label{eq:Dimerorder}
	\bra{\mathrm{MG}}d(t)\ket{\mathrm{MG}} =  \bra{\mathrm{MG}}U^{\dagger}(t, 0)~d~ U(t,0)\ket{\mathrm{MG}}  \\
	\an U(t,0) =\mathcal{T} \exp\Big( - \mathrm{i} \int_{0}^{t} \mathrm{d}t' H_{\mathrm{LR}}(t'; \alpha) \Big), \nonumber
\end{gather}
where $U(t, 0)$ denotes the usual time evolution operator from time $0$ to $t$ involving time ordering $\mathcal{T}$ in the Dyson sense. To simplify our notation we will suppress the powerlaw exponent $\alpha$ of the Hamiltonian from now on. Expanding expression \eqw{eq:Dimerorder} up to second order we get
\begin{align}
	\label{eq:short time expansion}
	d(t) &= \bra{\mathrm{MG}}d\ket{\mathrm{MG}} + \mathrm{i} \int_{0}^{t} \mathrm{d}t_{1} \bra{\mathrm{MG}} \comm{H(t_{1})}{d} \ket{\mathrm{MG}}  \nonumber \\
	&- \dfrac{1}{2} \int_{0}^{t} \mathrm{d}t_1 \mathrm{d}t_2
	\bra{\mathrm{MG}} \mathcal{T}\Big\{ \comm{H(t_1)}{\comm{H(t_2)}{d}} \Big\} \ket{\mathrm{MG}}.
\end{align}
Next we would like to comment on the terms of \eq{eq:short time expansion} individually. The first term describes just a static offset given by the maximal dimerization value $d_0 = 0.75$. The linear contribution in $t$ [second term in \eq{eq:short time expansion}] will vanish since the dimer order parameter only is summed up from two projectors on the singlet configuration $\ket{\mathrm{MG}}$ translated with respect to each other by one lattice site. This implies that $d \ket{\mathrm{MG}} \propto \ket{\mathrm{MG}} + \ket{\phi}$, where $\ket{\phi}$ is related to $\ket{\mathrm{MG}}$ by a projection onto the singlet configuration between two consecutive singlets of $\ket{\mathrm{MG}}$. As a consequence the linear part takes the form $\propto \operatorname{Im}[\bra{\mathrm{MG}}H(t_{1})\ket{\phi}]$, which in fact vanishes.\\
To simplify the last term, we choose  $t$ to be commensurate with full Floquet periods
\begin{gather}
	\label{eq:final form expansion}
	d(t) = d_0 - \dfrac{t^{2}}{2} \bra{\mathrm{MG}}\comm{H_{\mathrm{LR}}}{\comm{H_{\mathrm{LR}}}{d}}\ket{\mathrm{MG}}
\end{gather}
The last expression describes the short time evolution of the order parameter up to quadratic order and can be further simplified by evaluating the double commutator, see \fig{fig:Figure7-Appendix}{a}, which determines (the square of) the typical decay rate. 
The short-time expansion up to quadratic order is also compared to an exact numerical time evolution in \figc{fig:Figure7-Appendix}{b - d}.

\textit{System size dependence of thermal expectation values.---}We evaluate the system size dependence of the thermal phase diagram of \figc{fig:dynamics}{b} of the main text. In the main text, we already mentioned that computations for the finite temperature expectations of dimer order parameter and energy were performed based on the concept of typicality. Concretely we generate a number of $N$ random states, which will be evolved in imaginary time to $-\beta/2$. The entire set of $N$ evolved states can now be used as an approximation of the density matrix of the system.  Here, we evaluate the effective temperature after the quench for different system sizes ranging from 12 to 20 sites and different values of the long-range exponent  $\alpha\in\{0.2,0.5,1.0,1.5\}$. The results are illustrated in \fig{fig:Figure8-Appendix} for a summation over $N=40$ random states. We find good agreement with the data of \figc{fig:dynamics}{b} of the main text for all system sizes $L$. 

\textit{Numerical investigations of the dynamical Zak protocol.---}The protocol depicted in \figcc{fig:ZakPhase}{a}{b} of the main text can be used to measure topological phases in trapped ion experiments. The connection between this dynamical protocol and the conventional transformation of the Hamiltonian in Eq.~(5) of the main text is corroborated in this section. To this end, we simulate the protocol of \figcc{fig:ZakPhase}{a}{b} using exact diagonalization methods for systems of $L=15$ ions including the ancilla qubit. 

We start by implementing the initial state of our evolution protocol. We consider a system with an even number of sites, 
determine the ground state $\ket{\psi_{0}}$ of this system, and append an ancilla spin in state $\ket{\uparrow}_{A}$ to its left. The adiabatic ground state preparation is discussed in the next section. After performing a $\pi/2-$flip to the ancilla the system is described by the product state $\ket{\psi}\equiv (\ket{\uparrow}_{A} + \ket{\downarrow}_{A})/\sqrt{2} \otimes \ket{\psi_{0}}$. This state can now be evolved according to the dynamical protocol of \figcc{fig:ZakPhase}{a}{b}.

For sake of simplicity we will at first neglect the discretization of the original Floquet protocol within the time interval $[0, T_1]$ of a single period in our simulation. Instead we will directly use the effective description obtained in the high-frequency limit in terms of the long-ranged Heisenberg model $H_{\mathrm{LR}}$. Below, we will systematically study the full Floquet evolution. In this first part of the protocol, depicted in \figc{fig:ZakPhase}{a}, the ancilla qubit gets effectively decoupled from the remaining systems by $\pi-$pulses frequently applied to it. For the second part of the protocol within the time interval $[T_1, T_1+T_2]$ the decoupling of the ancilla no longer takes place. 
More precisely, the ancilla interacts with the remaining system via a long-ranged Ising Hamiltonian, whose orientation can be chosen to point along the $z$-direction. Interactions will furthermore exhibit a characteristic sign change in their prefactors as they cross the critical bond $i_c$. The evolution of the system during the second part of the protocol is hence described by the Hamiltonian
\begin{gather}
    H_{\mathrm{Zak}}(\alpha; i_c) = \sum_{0\leq i<j} J_{ij}(\alpha; i_c) S^{z}_{i} S^{z}_{j} \\
    \quad \text{with}\quad J_{ij}(\alpha; i_c)= \frac{3\sign(i_c - i)\sign(i_c - j)}{\vert j - i \vert^{\alpha}}.
\end{gather}
In this expression we identify the left most spin with the ancilla qubit $S^{z}_{0} \equiv \tau^{z}_{A}$. The critical bond $i_c$ in this notation takes half-integer values within $[1/2, L - 3/2]$. In every period of our Floquet protocol the evolution governed by $H_{\mathrm{Zak}}$ will induce an additional phase $\Delta\varphi(\alpha; i_c)$ to the Hamiltonian. Following the notation of the main text, the Hamiltonian will transform as
\begin{align}
    \label{eq:HamiltonianRotation}
    H_{\mathrm{LR}}(\alpha, \varphi) \longmapsto &\begin{cases}
      H_{\mathrm{LR}}(\alpha, \varphi + \Delta \varphi) & \text{for $\ket{\uparrow}_{A}$}\\
      H_{\mathrm{LR}}(\alpha, \varphi - \Delta \varphi) & \text{for $\ket{\downarrow}_{A}$}
    \end{cases}\\   
    \text{with} \quad \Delta\varphi(\alpha; i_c) &\equiv \Big(J_{0,i_c - \frac{1}{2}}(\alpha) -  J_{0,i_c +\frac{1}{2}}(\alpha)\Big) \frac{T_{2}}{2}\\
    &\equiv \Delta J(\alpha; i_c) \frac{T_2}{2}.
\end{align}
\begin{figure}
	\includegraphics[width=1.0\columnwidth]{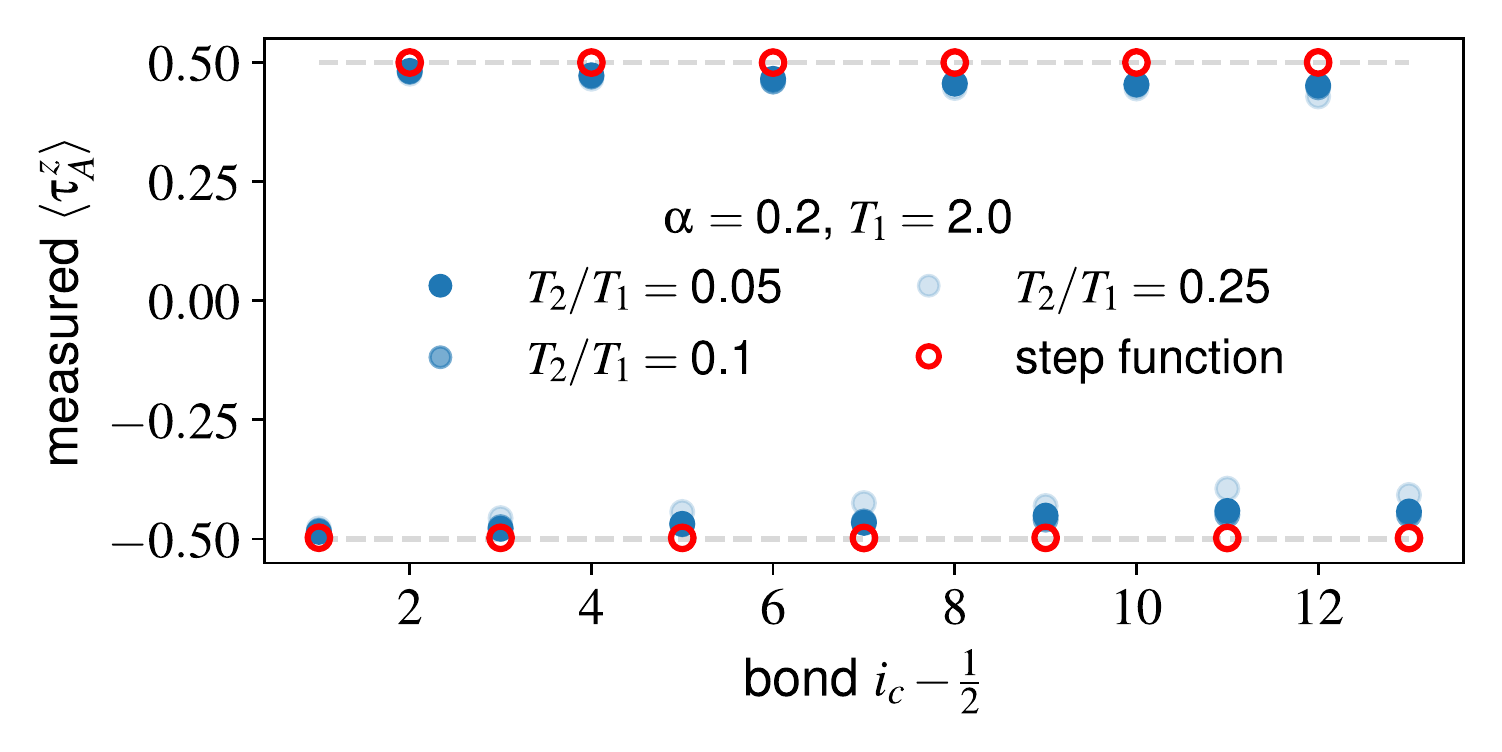}
	\caption{\textbf{Spatial dependence of Zak phases from the simulated protocol.} The phase accumulated at the ancilla $\exv{\tau^{x}_{A}}$ shows characteristic alternations between odd and even bonds. Imperfections of a power-law implementation of $H_{\mathrm{Zak}}$ become relevant at bonds further away from the ancilla qubit.
	}
	\label{fig:Figure11-Appendix}
\end{figure}
The factor of $\textstyle{\frac{1}{2}}$ is resulting from the fact that we use the spin operator of the ancilla $\tau^{z}_{A}$ in $H_{\mathrm{Zak}}$ instead of the related Pauli matrix. 

The absolute phase $\Delta\varphi(\alpha; i_c)$ acquired by the Hamiltonian thereby strongly depends on the value of the power-law exponent $\alpha$ as well as on the chosen critical bond $i_c$. Our implementation will describe an overall closed protocol if the relative phase between both ancilla states $\ket{\uparrow}_{A}$ and $\ket{\downarrow}_{A}$ equals $2\pi$.
When we start from the initial experimentally accessible Hamiltonian $H_{\mathrm{LR}}(\alpha, \varphi=0)$ we will find after $n$ periods Hamiltonians $H_{\mathrm{LR}}(\alpha, \varphi=\pm n\Delta\varphi)$ associated to the different $z-$basis states of the ancilla. This determines the number of Floquet periods $N$ needed for measuring the Zak phase as a function of $T_2, \alpha$ and $i_c$
\begin{equation}
    \label{eq:NumberFloquetPeriods}
    N = \frac{2\pi}{\Delta J(\alpha; i_c)T_{2}}.
\end{equation}
For large values of $\alpha$ and critical bond $i_c$, i.e. far away from the ancilla, $\Delta J(\alpha; i_c)$ can be very small, which requires either large application times $T_{2}$ or a high number of Floquet periods $N$ to successfully describe a closed protocol. We, however, find numerically that the number of Floquet periods needed for the considered systems stays reasonable small, as discussed below.

\begin{figure}
	\includegraphics[width=1.0\columnwidth]{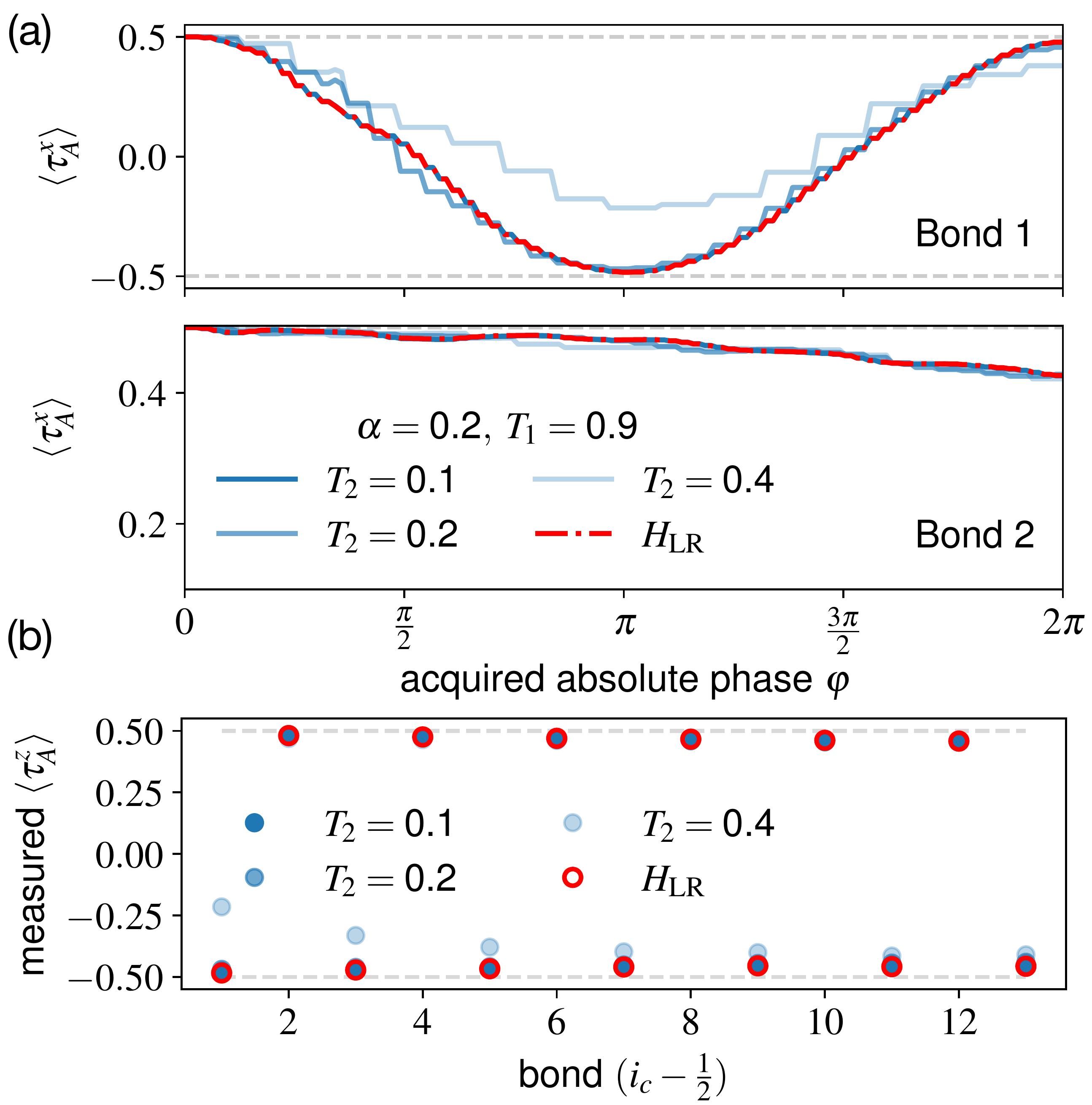}
	\caption{\textbf{Zak phase measurements for Floquet protocol.} (a) Change of the ancilla magnetization $\exv{\tau^{x}_{A}}$ during the evolution with a discretized version of the dynamical Zak protocol for $\alpha=0.2$ and application times for the sequence of Ising evolutions of $T_{1}=0.9$. The discretized protocol shows good agreement with the continuous evolution using $H_{\mathrm{LR}}$ (red dashed lines) for $T_{2}\lesssim 0.2$. (b) The results for interferometric measurements at $\varphi=\pi$ show the expected alternating behavior for different choices of $i_{c}$. 
	For large $T_{2}$ associated to a non-adiabatic implementation the protocol does not allow to reach $\exv{\tau^{z}_{A}}=-\textstyle{\frac{1}{2}}$ at odd bonds. However, the alternating nature of the pattern can be resolved even away from the adiabatic limit $T_{2}\ll T_{1}$. 
	}
	\label{fig:Figure12-Appendix}
\end{figure}

Before investigating detailed results of numerical simulations, we want to address another aspect important for measurements of topological phases at the ancilla qubit. Besides the desired effect of transferring the topological Zak phase to the ancilla spin, we also need to study the consequences of our protocol of \figc{fig:ZakPhase}{a} on the physical spin system.  To this end, we interpret the evolution governed by $H_{\mathrm{Zak}}$ as a rotation for each spin with specific angle $\delta\varphi$ around the $z$-direction, $R_{\delta \varphi}\equiv \exp(-\iu T_{2}H_{\mathrm{Zak}})$. 
To simplify our argument, we will for the moment consider a coupling in the form of  step function ($\alpha \to 0$ case). Spins located left/right to the critical bond $i_{c}$ will thereby acquire rotations with opposite chirality. Acting with these rotations on our Hamiltonian, i.e., computing $R_{\delta\varphi}^{\dagger}H_{\mathrm{LR}}(\alpha, \varphi)R_{\delta\varphi}^{\phantom{\dagger}}$, exactly reproduces the transformation of \eq{eq:HamiltonianRotation}. The rotation $R_{\delta\varphi}$ hence allows us to describe the evolution with all Hamiltonians within the family $\{H_{\mathrm{LR}}(\alpha,\varphi_{n})\}_{n}$, although not directly accessible in experiments. In total, we find for the evolution in our closed protocol of \figc{fig:ZakPhase}{b},
\begin{align}
    \label{eq:ProtocolDisc}
    \prod_{n=1}^{N} \Big(\eu^{-\iu T_{1} H_{\mathrm{LR}}(\alpha,\varphi_{n})}\Big) &= \prod_{n=1}^{N} \Big( \big( R^{\dagger}_{\delta \varphi}\big) ^{n}\eu^{-\iu T_{1} H_{\mathrm{LR}}(\alpha, 0)}\big(R^{\phantom{\dagger}}_{\delta \varphi}\big)^{n}\Big) \nonumber\\
    &= \big( R^{\dagger}_{\delta \varphi}\big) ^{N}\prod_{n=1}^{N} \Big(\eu^{-\iu T_{1} H_{\mathrm{LR}}(\alpha, 0)}R^{\phantom{\dagger}}_{\delta \varphi}\Big), 
\end{align}
where $\{\varphi_{n}\}_{n}$ denotes a suitable discretization of the interval $[0, 2\pi]$. The second factor of the last equality is precisely the evolution described by steps (ii) and (iii) in the protocol of \figc{fig:ZakPhase}{b} in the main text. The first factor, moreover, indicates an inverse rotation around an angle $N\delta\varphi$, which also has to be applied in our protocol before measuring. This inverse rotation takes care of an additional unwanted phase acquired during the protocol evolution in the physical spin system. Neglecting this contribution would result in interferometric measurements not only detecting the desired topological phase at the ancilla, but also an unknown phase acquired in the system.  
The required inverse rotation of \eq{eq:ProtocolDisc} is obtained by applying $-H_{\mathrm{Zak}}$ for the same duration as the overall application time of $H_{\mathrm{Zak}}$ during the protocol. The Hamiltonian $-H_{\mathrm{Zak}}$ can thereby be generated analogously to its positive counterpart $H_{\mathrm{Zak}}$. Specifically, after the last Floquet period of the protocol, a single global $\pi$-pulse to the left part of the system (in contrast $H_{\mathrm{Zak}}$ was generated by a $\pi$-pulse acting on the right part) implements $-H_{\mathrm{Zak}}$ and subsequent time evolution finally allows to
unwind the unwanted phase accumulation of the physical spin system. This subtlety is taken into account in step (iv) of the protocol of \figc{fig:ZakPhase}{b}. 
After this procedure the Zak phase of the ancilla can be measured in the conventional way on the ancilla qubit.

With the basics of our simulations set, let us have a look at concrete results. 
In \fig{fig:Figure10-Appendix} we provide results for dynamical evolution of the initial state $\ket{\psi}$ for different choices of $T_{2}/T_{1}$ and the critical bond $i_{c}$ in a system with power-law scaling $\alpha=0.2$. We show the magnetization of the ancilla qubit in $x-$basis $\exv{\tau^{x}_{A}}$ against the absolute phase $\varphi$ acquired by the Hamiltonian $H_{\mathrm{LR}}(\alpha, \varphi)$. The interferometric measurement to extract the value of the Zak phase should be performed at $\varphi=\pi$ (orange dashed lines), where both absolute phases add up to a relative phase difference of $2\pi$. We can clearly identify odd bonds (top row: bond 1, bond 3, ...) and even bonds (bottom row: bond 2, bond 4, ...) in \fig{fig:Figure10-Appendix}. Whereas odd bonds approach the value $\exv{\tau^{x}_{A}} = -\textstyle{\frac{1}{2}}$ for $\varphi\to\pi$, even bonds will almost return to the starting value $\exv{\tau^{x}_{A}} = +\textstyle{\frac{1}{2}}$. This refers to values of the Zak phase of $\gamma_{\mathrm{Zak}}=\pi$ and $\gamma_{\mathrm{Zak}}=0$, respectively, considering an ancilla state of
\begin{equation}
    \ket{\psi(\varphi=\pi)}_{A} = \frac{1}{\sqrt{2}}\big( \ket{\uparrow} + \eu^{\iu \gamma_{\mathrm{Zak}}}\ket{\downarrow} \big)
\end{equation}
at absolute implemented phase of $\varphi=\pi$. To investigate the impact of a power-law decay in the couplings of $H_{\mathrm{Zak}}$ we furthermore show in \fig{fig:Figure10-Appendix} results obtained using a perfect step function ($\alpha\to0$) in $H_{\mathrm{Zak}}$ (red dashed lines). We find that deviations between power-law and step function are most pronounced for even bonds: Whereas the step function implementation returns to the starting value at absolute phases of $\varphi=\pi$, the power-law results yield values of $\exv{\tau^{x}_{A}}<\frac{1}{2}$. This imperfection is, however, expected as a power-law entails small but finite values for the implemented phases also away from the critical bond $i_c$. As a consequence our protocol will apart from the chosen critical bond $i_c$ also effect couplings across other bonds; most dominantly the first one. These phases implemented at other bonds will in general prohibit recurrence to the initial configuration at $\varphi=\pi$. This also applies in similar form for a phase accumulation of $\varphi=2\pi$, shown in \fig{fig:Figure10-Appendix}. At $\varphi=2\pi$, where we expect our protocol to reproduce the initial state independent of the choice for $i_c$, only the step function implementation yields fully accurate results. Despite the power-law decay of interactions causing some deviations from the ideal behavior, we find that the protocol of \figcc{fig:ZakPhase}{a}{b} allows clear discrimination of even and odd bonds, respectively. This becomes clear when extracting the values of $\exv{\tau^{z}_{A}}$ interferometrically measured at the ancilla for $\varphi=\pi$ (orange dashed lines) as a function of the chosen bonds $i_{c}$, see \fig{fig:Figure11-Appendix}.

Now, we can repeat the analysis for the evolution under the discretized Floquet protocol of consecutive rotated long-ranged Ising Hamiltonians. We proceed in exactly the same manner as before and expect the results to be qualitatively equivalent if we consider the adiabatic limit $T_{2}\ll T_{1}$ with the drive frequency of the Floquet protocol $\propto 1/T_{1}$ still being sufficiently fast. We show the results for the discretized version of the dynamical protocol in \fig{fig:Figure12-Appendix}. Indeed we find that for example for a choice of $(\alpha, T_{1})=(0.2, 0.9)$ values of $T_{2}\lesssim 0.2$ are already sufficient to obtain the correct behavior we already encountered for evolution governed by the high-frequency Hamiltonian $H_{\mathrm{LR}}$ (red dashed lines). This is illustrated in \figc{fig:Figure12-Appendix}{a} for evolution associated to $i_c$ located at the first two bonds in the system (continuous data obtained for $T_{1}=0.9$ and $T_{2}=0.1$).
Extracting the results for the interferometric measurement for all bonds indicates that also larger $T_{2}\approx0.4$ already yield the characteristic alternating pattern of the Zak phase in the topological phase, as illustrated in \figc{fig:Figure12-Appendix}{b}.

Given all this input, we can estimate the number of Floquet periods needed to obtain the results of \fig{fig:Figure10-Appendix}, \fig{fig:Figure11-Appendix} and \fig{fig:Figure12-Appendix}. According to \eq{eq:NumberFloquetPeriods} the number of Floquet periods $N$ strongly depends on $\alpha$ and the chosen $i_{c}$. For our simulations this requires for example $N\le 9$ Floquet periods to reach the desired absolute phase of $\varphi=\pi$ for $(\alpha, T_{2})=(0.2, 0.2)$ and $(i_{c}-\textstyle{\frac{1}{2}})\le13$. For larger values of $\alpha$~(e.g. $\alpha=0.8$), the required number of periods can vary more drastically with $i_{c}$~($N=7$ for $i_{c}=1$ or $N=40$ for $i_{c}=12$). Overall, the requirements are, however, not too severe highlighting the experimental relevance of the proposed protocol. 
\begin{figure}
	\includegraphics[width=1.\columnwidth]{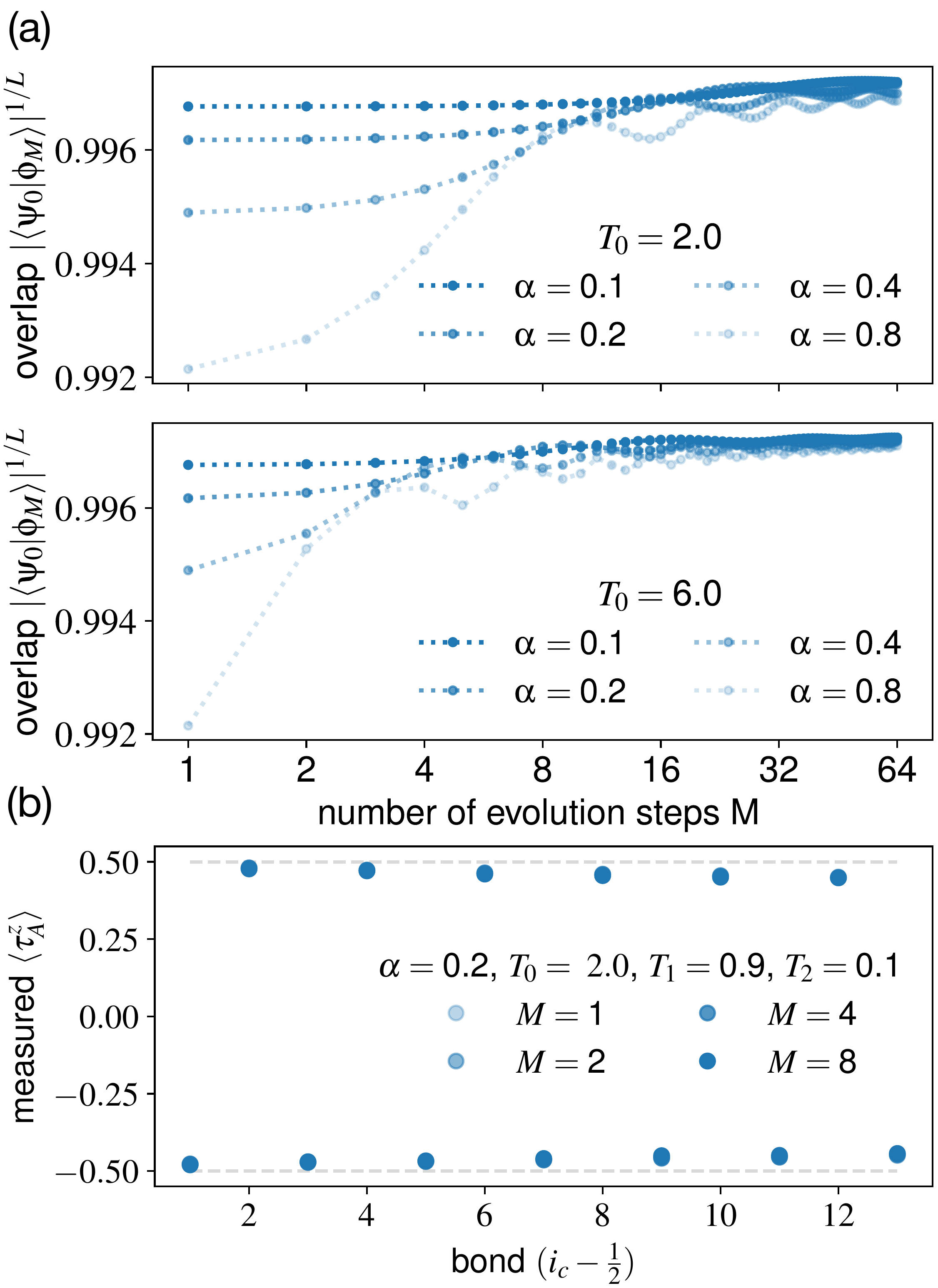}
	\caption{\textbf{Adiabatic ground state preparation.} (a) We show the $L-$th root of the absolute overlap between an in $M-$steps adiabatically prepared state $\ket{\phi_{M}}$ and the ground state $\ket{\psi_{0}}$ of $H_{\mathrm{LR}}(\alpha)$ for different values of the power-law scaling $\alpha$ and application times of $T_{0}\in\{2.0, 6.0\}$ for each evolution step. We find that already a comparable small number of evolution steps $M$ is sufficient to reach large overlap with the actual ground state for small $\alpha\lesssim0.8$. (b) Using the adiabatically prepared states as initial states for our dynamical protocol we find qualitatively similar results as for the previous simulations.}
	\label{fig:Figure13-Appendix}
\end{figure}

\textit{Adiabatic ground state preparation.---} In the previous section we assumed that the system can initially be prepared in its ground state $\ket{\psi_{0}}$. This can for example be achieved by adiabatic evolution. To this end, we start out with a fully dimerized state $\ket{\mathrm{MG}}$. To approximate the ground state of $H_{\mathrm{LR}}(\alpha)$ we sequentially evolve $\ket{\mathrm{MG}}$ with $H_{\mathrm{LR}}(\alpha_{m})$ for a given time interval $T_{0}$. The values of $\alpha_{m}$ thereby approach the final value of $\alpha$ in $M$ steps, i.e. $\alpha_{m}\equiv m\;\delta\alpha$ for $m\in\{1, 2, ,\dots, M\}$ with $\delta\alpha=\textstyle{\frac{\alpha}{M}}$. If the resulting state $\ket{\phi_{M}}$ is sufficiently close to the ground state of the system $\ket{\psi_{0}}$ it should allow for measurements of Zak phases with results similar to the ones displayed in \figg{fig:Figure11-Appendix}{fig:Figure12-Appendix}.

We show the $L-$th root of the absolute overlap $\vert\bracket{\psi_{0}}{\phi_{M}}\vert^{1/L}$ between the adiabatically prepared state and the true ground state in \figc{fig:Figure13-Appendix}{a} for various choices of $\alpha$ as a function of the number of evolution steps $M$. During the evolution every Hamiltonian $H_{\mathrm{LR}}(\alpha_{m})$ is applied for a time interval of $T_{0}=2.0$ respectively $T_{0}=6.0$. We notice that already a comparable small number of evolution steps $M$ provides a good approximation of the actual ground state for small values of $\alpha\lesssim 0.8$. We moreover repeat the simulations for the interferometric protocol starting from these states. Our results, shown in \figc{fig:Figure13-Appendix}{b}, reveal that adiabatically prepared initial states yield similarly to previous results an alternating pattern in the interferometrically measured magnetization $\exv{\tau^{z}_{A}}$ at the ancilla, as expected from the good fidelity of the adiabatically prepared ground state.

Combining all insight we gained within the last paragraphs we can estimate the experimental requirements for interferometric measurements of topological Zak phases. Once more concrete statements highly depend on the chosen parameters. We will in the following hence consider exemplary choices reasonable close to experimental possibilities. Recalling the $T_{2}-$dependence of \eq{eq:NumberFloquetPeriods} we want to choose the application time of $H_{\mathrm{Zak}}$ as large as possible, while still remaining close enough to both the adiabatic and the high-frequency limit. Our simulations of the discretized protocol indicate that this criterion is e.g. satisfied for $(T_1, T_2)=(1.2, 0.4)$. Assuming initial ground state preparation within $M=8$ evolution steps we expect $\gamma_{\mathrm{Zak}}$
to be measurable with $50$ global and $32$ local pulses at the first two bonds of a system characterized by a power-law exponent $\alpha=0.8$. This estimate includes $1$ local and $3M$ global pulses for preparation of the initial state, $6N$ local respectively $5N$ global pulses for evolution as well as $1$ additional global and local pulse for unrotating the system and performing the interferometric measurement.

\bibliography{LRHeisenberg}


\end{document}